\documentclass[letterpaper,twocolumn,10pt]{article}
\usepackage{SOUPS2021}
\usepackage{tikz}
\usepackage{amsmath}
\usepackage{filecontents}
\usepackage{dirtytalk}
\usepackage{tabularx}
\usepackage{lscape}
\usepackage{arydshln}
\usepackage{makecell}
\usepackage{mdframed}
\usepackage{pbox}
\usepackage[most]{tcolorbox}
\usepackage{booktabs, subcaption, amsfonts, dcolumn}
\usepackage{multirow}
\usepackage{todonotes}
\usepackage{comment}
\usepackage{colortbl}
\usepackage{makecell}
\usepackage{hhline}
\usepackage{todonotes}
\usepackage{enumitem}
\usepackage{framed}
\usepackage{soul}
\usepackage[usestackEOL]{stackengine}
\definecolor{shadecolor}{HTML}{D1E8E2}
\mdfdefinestyle{MyShadeQuoteStyle}{
    backgroundcolor=shadecolor,
    linewidth=0pt,
    skipbelow=\topskip,
    skipabove=30pt,
    usetwoside=false,
    innertopmargin=10pt,
    innerbottommargin=10pt,
    innerleftmargin=0.5cm,
    innerrightmargin=0.5cm,
    roundcorner=10pt
}

\newcommand{\boldpartitle}[1]{\vspace{.1in}\noindent\textbf{#1.}}

\usepackage{titlesec}
\titlespacing*{\section}{0pt}{\baselineskip}{.4\baselineskip}
\titlespacing*{\subsection}{0pt}{.7\baselineskip}{.3\baselineskip}
\titlespacing*{\subsubsection}{0pt}{.5\baselineskip}{.1\baselineskip}

\definecolor{note_blue}{HTML}{035aa6}
\definecolor{note_green}{HTML}{158467}
\definecolor{note_red}{HTML}{d54062}
\definecolor{Gray}{gray}{0.9}
\definecolor{Mahogany}{HTML}{C04000}
\definecolor{RoyalBlue}{HTML}{0066CC}

\usepackage{graphicx,amssymb}
\newcommand{\widesim}[2][1.5]{
  \mathrel{\overset{#2}{\scalebox{#1}[1]{$\sim$}}}
}

\hypersetup{colorlinks,urlcolor=RoyalBlue}
\hypersetup{colorlinks,citecolor=Mahogany}
\hypersetup{linkcolor=Mahogany}

\begin{document}

\date{}

\title{\Large \bf Understanding Privacy Attitudes and Concerns Towards Remote Communications During the COVID-19 Pandemic}

\def\plainauthor{Pardis Emami-Naeini, Tiona Francisco, Tadayoshi Kohno, Franziska Roesner}

\author{
{\rm Pardis Emami-Naeini, Tiona Francisco, Tadayoshi Kohno, Franziska Roesner}\\
Paul G. Allen School of Computer Science \& Engineering \\
University of Washington
}

\maketitle
\thecopyright

\begin{abstract}
Since December 2019, the COVID-19 pandemic has caused people around the world to exercise social distancing, which has led to an abrupt rise in the adoption of remote communications for working, socializing, and learning from home. As remote communications will outlast the pandemic, it is crucial to protect users' security and respect their privacy in this unprecedented setting, and that requires a thorough understanding of their behaviors, attitudes, and concerns toward various aspects of remote communications. To this end, we conducted an online study with 220 worldwide Prolific participants. We found that privacy and security are among the most frequently mentioned factors impacting participants' attitude and comfort level with conferencing tools and meeting locations. Open-ended responses revealed that most participants lacked autonomy when choosing conferencing tools or using microphone/webcam in their remote meetings, which in several cases contradicted their personal privacy and security preferences. Based on our findings, we distill several recommendations on how employers, educators, and tool developers can inform and empower users to make privacy-protective decisions when engaging in remote communications.
\end{abstract}

\section{Introduction} \label{sec:intro} 
The world was hit by a pandemic caused by the novel coronavirus and the COVID-19 disease in December 2019. In an attempt to prevent the spread of the virus, businesses and schools around the globe shut down, and people began sheltering in their homes to practice social or physical distancing~\cite{social-world}.

Following social distancing protocols, people around the world have been encouraged or ordered to stay at home~\cite{uk-home, china-home, us-home}. Hence, they started to work from home~\cite{wfh}, keep in touch with family and friends remotely~\cite{socializing}, and/or take remote courses~\cite{education}, many for the first time~\cite{remote-edu-first}. This made remote conferencing tools an essential part of people's day-to-day lives, which albeit useful, posed potential privacy risks to people who are now regularly streaming video and audio from their own homes~\cite{flaw1, messenger-risk, skype-risk, whatsapp-vul7, zoom-vul3, work-risk-1, edu-risk-1}. 

Increasingly integrated into people's lives and routines, widespread remote communications will not disappear with the end of the pandemic~\cite{after-survey}. As people continue to work, socialize, and learn from home, it becomes imperative for their privacy and security to be protected. This requires the designers of in-home technologies (e.g., conferencing tools) and organizations that use them to understand the diverse needs of users. Understanding users' needs will enable designers and organizations to i) inform users about potential risks, and ii) gear their designs toward enabling users to control their privacy and security when using such technologies.

To that end, we conducted a worldwide survey ($n=$220) on Prolific~\cite{prolific} in May 2020, i.e., a few months into the pandemic, as people were newly settling into widespread remote communications. We sought to conduct our study during the transition phase of the COVID-19 pandemic, when participants were still adjusting to their new remote settings, while remembering their normal lives before the pandemic. Our survey covered three contexts of remote communications, namely work from home (WFH), socialize from home (SFH), and learn from home (LFH). In each context, without priming participants by asking directly about privacy and security, we leveraged participants' open-ended responses to tease out their unbiased privacy and security attitudes and behaviors towards three aspects of remote communications: conferencing tools, modes of remote communications (microphone, webcam), and locations of remote communications. We conducted quantitative and qualitative analyses to answer the following three research questions:

\begin{enumerate}[itemsep=0mm]
    \item How do people engage with different aspects of remote communications in each context during the pandemic?
    \item How do privacy and security factor into people's behaviors and attitudes towards aspects of remote communications in each context?
    \item What approaches can be used to effectively inform and empower users' privacy and security decision making related to remote communications in each context?
\end{enumerate}

We also designed our survey to allow us to explore related research questions for other technologies that we hypothesized that people would interact with more and/or have a new relationship with during the pandemic stay-at-home orders: smart home devices and social media platforms. Upon analyzing our results, we found that privacy and security concerns of most participants have not changed during the pandemic, mainly due to their lack of risk knowledge and awareness. Stay-at-home orders, however, significantly impacted participants' concerns and behaviors toward remote communications, which we primarily focus on in this paper. For completeness, we include the survey questions on smart home devices and social media platforms in Appendix~\ref{survey_questions} and a summary of their findings in Appendices~\ref{smart-home} and~\ref{social-media}.

When being asked about conferencing tools, participants expressed a lack of decision-making agency. In WFH and LFH, participants reported to use the tool that was being decided for them by their employer or educator. Moreover, in all contexts, participants felt that they had no control over activating their webcam/microphone during their remote communications. For several participants, such imposed requirements contradicted their privacy and security preferences.  

We found that participants' privacy attitudes and concerns towards the physical locations where their remote communications take place is context-dependent. By qualitatively analyzing participants' open-ended responses, we identified two types of location-related privacy: remote privacy (privacy from meeting attendees) and co-inhabitant privacy (privacy from household members). The open-ended responses suggested that in SFH, participants are mainly concerned about their co-inhabitant privacy, while valuing both remote and co-inhabitant privacy in WFH and LFH.

Based on the outcomes of our study, we distill several recommendations for organizations and tool developers on how to more effectively enable users to make informed and privacy-protective decisions with regard to their remote communications. In particular, we propose to enhance users' decision-making process by means of inclusive, transparent, and flexible policies on remote communications and designing privacy-protective features, which consider diverse and context-specific privacy and security needs.
\section{Background and Related Work}
Since December 2019, people around the world have been struggling with SARS-CoV-2 (novel coronavirus) and the resulting COVID-19 pandemic~\cite{covid-struggle}. To help prevent further spread of the virus, many people have exercised social or physical distancing, i.e., keeping a safe distance from others who are not from the same household~\cite{CDC_social-distancing}. Consequently, people started working, socializing, and learning from home. As a result, the use of conferencing tools and audio and video communications has increased dramatically. 

\subsection{Privacy and Security Risks of Conferencing Tools}
The pandemic has redefined home from a place of \textit{privacy and security}~\cite{hayward1975home} to a shared work, socializing, and learning space. This sudden shift from in-person to remote interactions has led to an unprecedented increase in the use of remote communication tools~\cite{surge-apps}. Teleconferencing and video conferencing tools, such as Zoom~\cite{zoom}, Microsoft Teams~\cite{ms-teams}, Google Hangouts~\cite{google-hangout}, and WebEx~\cite{cisco}, have all seen a massive rise in usage thanks to people working, socializing, and learning from home.

As people started to increasingly rely on such tools for their daily communications, experts have become more concerned about the wide range of privacy and security risks these tools expose their users to~\cite{flaw1, skype-risk, whatsapp-vul7, messenger-risk}. A few of the reported concerns include Zoombombing~\cite{zoom-vul2}, undisclosed data mining~\cite{zoom-vul5}, and selling information to third parties~\cite{zoom-vul4}. By considering the context around remote communications, literature has discussed the privacy and security concerns involving remote health-related sessions~\cite{li2020privacy, bassan2020data}, educational communications~\cite{edu-risk-1, edu-risk-2, edu-risk-3, edu-risk-4}, attending online courses~\cite{lfh-concern1, lfh-concern2}, and work-related meetings~\cite{work-risk-1, work-risk-2, work-risk-3}.

Experts have provided several guidelines aiming to prevent the risks and mitigate the potential harms of conferencing tools~\cite{hygiene, digital-privacy, protect1}. Despite being valuable sources of information, these guidelines put the burden of protecting privacy and security mainly on the user. This is an unrealistic expectation due to several reasons. Confirming the literature~\cite{emami2019exploring}, our findings showed that privacy and security aspects, although being important, are not always the number one priority when using and interacting with conferencing tools. Moreover, our qualitative findings suggested that due to their roles in their organizations, users often have limited power in making privacy-protective decisions, especially in work- and education-related contexts. In addition, the best practices reported in the current guidelines constitute a broad recipe, hoping that they apply to all users in all contexts of remote communications. From the literature, we already know that privacy is context-dependent~\cite{naeini2017privacy}. 

\subsection{Home Audio and Video Broadcasting}
During the pandemic, people started to rely more and more on the microphones and webcams of their devices to stay in touch with their colleagues, friends and family members, or their classmates. Only a few weeks into the pandemic, the market saw a 179\% jump in the sales of webcams~\cite{webcam-sale}, followed by a supply shortage~\cite{webcam-sold-out-1, webcam-sold-out-2, webcam-sold-out-3}. 

Privacy and security experts have indicated that webcams and microphones are susceptible to risks and vulnerabilities. Several reports showed how easily hackers take control of users' devices and activate their built-in webcam and microphone by exploiting the device vulnerabilities~\cite{hack1, hack2, hack3, hack4, hack5, hack6}. During the pandemic, in all contexts of remote communications, users are at an even higher risk of such hacking incidents as they are spending an increased amount of time using their webcams and microphones in different locations of their homes to remotely communicate with others~\cite{risk-covid}. Users might not be aware that their webcams and microphones are turned on as the LED indicator lights are not always effective~\cite{portnoff2015somebody} or they might have been deactivated by the attacker~\cite{brocker2014iseeyou, indicator1}.

To prevent hacking attacks from happening, experts frequently recommend users to cover their webcams and microphones when they are not being used~\cite{webcam-mic-protect1, webcam-mic-protect2}. During the pandemic, however, users might not be able to diligently exercise this protective approach as many are encouraged or even forced to have their webcams and microphones on all the time. Employers are setting always-on webcam policies to encourage spontaneous chats among employees~\cite{worker3} and using surveillance tools to closely monitor the activities of their workforce~\cite{worker1, worker2}, in some cases even without users' knowledge~\cite{worker4}. Saying no to such surveillance is not always easy, especially during the pandemic with the heightened risk of unemployment due to potential retaliations~\cite{unemployed}.

Remote learning is not immune to such commonplace imposed surveillance as well. School-issued devices are not transparent about whether they spy on students by activating their webcams and microphones~\cite{student1}. Some schools use proctoring software that enables access to the students' webcams and microphones during the exams~\cite{student3, proctor}. In addition, policies are in place forcing students to have daily audio and video interactions with their peers or teachers~\cite{student2}.

The aforementioned privacy-invasive webcam and microphone policies and surveillance practices allude to the ineffectiveness of the blanket and commonly referenced solutions with respect to the rising risks of these technologies. Designing privacy-protective tools and providing usable privacy and security guidelines for users require a deep understanding of users' decisions and behaviors. Our study contributes to the body of literature by providing novel empirical evidence, which highlights the significant impact of the context of remote communications, as well as the living conditions, on attitudes and privacy concerns related to remote communications.
\section{Methods} \label{sec:method}
We launched an online worldwide survey ($n=$220) on Prolific in May 2020. We initially recruited 230 participants and excluded 10 of them: 3 participants used the open-ended boxes to advertise a product and 7 participants provided other irrelevant responses to open-ended questions. We provide the complete list of survey questions in Appendix~\ref{survey_questions}, and we mention the question number in parenthesis (e.g., CQ1) when referring to each survey question in the remainder of this section. The study protocol was approved by our Institutional Review Board (IRB).

\subsection{Participant Recruitment}
Prior to the survey launch, we piloted our survey with 10 participants to identify potential issues with the survey questions, specify the compensation amount, and ensure that the questions are all understandable and easy to follow. We then conducted our survey on Prolific and recruited participants who were at least 18 years old. Because of the worldwide impact of the COVID-19 pandemic, we did not restrict our respondents to a specific region and instead, recruited participants from all around the world. The survey took on average 16 minutes to be completed, and we compensated each participant with US\$5.

\subsection{Survey Procedure}
We started the survey by introducing our study to be about \say{technology use in the home during the Coronavirus (COVID-19) Pandemic.} We then asked a few questions to obtain participants' consent to participate in our study (see Appendix~\ref{consent}).

We asked questions on three contexts of remote communications: working from home (WFH), socializing from home (SFH), and learning from home (LFH). In the survey, we showed questions related to each context in a separate block. We randomized the order of these blocks to mitigate the potential order bias~\cite{perreault1975controlling}. We asked similar questions in the three tested contexts and only changed how we referred to remote communications in each context. Specifically, in the contexts of WFH, SFH, and LFH, we referred to remote communications as \say{remote work-related meetings,} \say{remote personal meetings with friends and family members,} and \say{remote learning-related meetings,} respectively.

\subsubsection{Context-Specific Questions}
Familiarity with a situation has been shown to impact people's risk perception toward that situation~\cite{weber2005communicating}. To control for participants' familiarity with the contexts of remote communications, at the beginning of each context, we asked participants to specify whether they have experience with remote communications in that context (CQ1). We implemented a logic so that respondents could see the remaining questions of that context only if they reported to have experience with the context in question.

To better understand our participants' timeline for remote communications, we asked questions to capture when they started remote communications and how often they were engaged in remote communications before and during the pandemic (CQ2-4). In each context, we explored participants' attitudes, behaviors, and privacy concerns related to three aspects of remote communications: conferencing tools (CQ5-10), modes of remote communications (CQ11-14), and locations of remote communications (CQ15-19). 

There are countless anecdotes that have been reported about people's remote communication experience during the pandemic~\cite{story-learning}. To understand what our participants were most concerned about in their remote communications, at the end of each context, we asked respondents to specify the incidents that happened to themselves or others that they perceived to be concerning or awkward (CQ20-22).

\subsubsection{Smart Home Devices and Social Media Platforms}
After context-specific questions, we asked a few questions to explore how the pandemic impacted participants' privacy concerns and behaviors toward their smart home devices and social media platforms (IQ1-9, SMQ1-8). Since these questions were not the main focus of our study, we provide their findings along with a background in Appendices~\ref{smart-home} and~\ref{social-media}.

\subsubsection{Demographics and Home Settings}
Finally, we asked questions to understand participants' demographic information, as well as their home settings (DH1-16). We placed the demographic questions at the end of the survey to minimize the possibility of stereotype threat~\cite{fernandez2016more, steele1997threat, spencer1999stereotype}.

\subsection{Data Analysis}
To analyze participants' responses to survey questions, we conducted qualitative and quantitative analyses.

\subsubsection{Qualitative Analysis}
The first author was the primary coder, who created the codebook for each open-ended question and kept it updated throughout the coding process. To analyze the data, we applied structural coding~\cite{saldana2015coding}, which is a question-driven qualitative coding approach to categorize the interview data as well as open-ended survey responses~\cite{namey2008data}. The codebook consists of main and sub codes. The main codes are created from the topics of interest in the study. For example, we were interested in understanding what factors led participants to use specific conferencing tools during the pandemic. In the codebook, the main code we used to answer this research question was \textit{reasons to use conferencing tools}, which was then divided into 11 sub-codes (e.g., \textit{functionality}) and further divided into 8 sub-sub-codes (e.g., \textit{convenience and accessibility}). After the codebook was created, the first two authors used the codebook to independently code all the open-ended responses. Authors had several meetings to go over the codebook and the coded responses and resolve the conflicts stemming from mismatched understandings of the codebook. After agreeing on the definitions used in the codebook, the first two authors re-coded all the responses. The final codebook consists of 11 main codes, 122 sub-codes, 54 sub-sub-codes, and 4 sub-sub-sub-codes. For each codebook, the Cohen's Kappa inter-coder agreement was calculated after the second round of coding. The average rate of agreement for all the codebooks was above 91\%, with a minimum of 88\% and a maximum of 100\%. Based on the literature, Cohen's Kappa inter-coder agreement of over 75\% is considered as \say{excellent}~\cite{fleiss2013statistical}. We provide the final codebooks in Appendix~\ref{codebooks}.

\subsubsection{Quantitative Analysis}
For our regression analysis, we fit $M=4$ Cumulative Link Mixed Models (CLMMs) with logit as the link function to our collected data in order to explain the dependent variables (DVs) we asked our participants about. In each model, the DV is a categorical variable that can take multiple \emph{ordinal} values, each of which we refer to as a \emph{response category}. For all models, we treated participants' demographic and home setting factors as control variables. We considered Akaike Information Criterion (AIC) as the goodness of fit for the models~\cite{burnham2004multimodel}. We only report on demographic factors that helped the model fit significantly better than the model without them. It is important to note that we did not include the interaction terms in the final regression models as they did not improve the model fit. For the $m$\textsuperscript{th} model, $m\in\{1,\dots,M\}$, we denote the number of possible response categories by $J_m$, and we denote the corresponding number of \emph{observations}, i.e., the number of participants that answered the question corresponding to that model, by $N_m$. For the $n$\textsuperscript{th} observation, $n\in\{1,\dots,N_m\}$, we let $Y_m^n$ denote the observed response category. As per the CLMM definition, for the $m$\textsuperscript{th} model, $m\in\{1,\dots,M\}$, the probability that the $n$\textsuperscript{th} observation, $n\in\{1,\dots,N_m\}$, falls in the $j$\textsuperscript{th} response category or below, $j\in\{1,\dots,J_m-1\}$, is modeled as
\begin{align*} 
\text{logit}(\Pr(Y_m^n\leq j)) 
&= \alpha_{j|j+1} - u_{\texttt{participant}_n} - \sum_{i=1}^{I_m} \beta^n_{\texttt{IV}_{m,i}}, \label{eq:clmm_model}
\end{align*}
where $\alpha_{j|j+1}$ denotes the threshold parameter or cut-point between response categories $j$ and $j+1$, and $u_{\texttt{participant}_n}\widesim[2]{i.i.d.} \mathcal{N}(0, \sigma_u^2)$ denotes the random effect for the participant in the $n$\textsuperscript{th} observation. Moreover, $\{\beta^n_{\texttt{IV}_{m,i}}\}_{i=1}^{I_m}$ represent model coefficients corresponding to the $I_m$ different independent variables (IVs) in the $m$\textsuperscript{th} model, each particular to the level that was reported in the $n$\textsuperscript{th} observation.
\section{Results} \label{sec:results}
We start this section by providing information on participants' demographics and timelines of remote communications. We then present findings on participants' behaviors and decisions related to three aspects of remote communications: conferencing tools, modes of remote communications, and locations of remote communications.

\subsection{Participants}
We recruited 230 participants (reduced to 220 after excluding invalid responses) on Prolific. Our participants were mainly from UK (31\%), Poland (15\%), and US (14\%). 43\% of our respondents were female and 57\% were male. Most participants did not have a background in Information Technology fields (65\%) and were 18-29 years old (62\%). We provide details on participants' demographics, home settings, and timelines of remote communications in Appendix~\ref{demographics}. Except for questions on the consent form (see Appendix~\ref{consent}), none of the survey questions required participants to provide an answer. For each finding, we specify the number of participants that answered the corresponding question.

\subsubsection{Frequency of Remote Communications}
When asked about the frequency of remote communications before the pandemic, responses suggested that participants had more experience with remote communications in the socializing context than work and learning contexts. During the pandemic, in the contexts of WFH and SFH, most participants (WFH:~150/220, SFH:~208/220) reported that they have been mostly having remote meetings and communications. In the context of LFH, about half of our respondents (LFH:~114/220) reported to be having remote learning-related meetings.

Designing usable and privacy- and security-protective conferencing tools and guidelines in remote communications requires a deep understanding of users' attitudes and concerns towards remote communications. To this end, in our survey, we captured participants' context-specific thought process and decision making toward three aspects of remote communications during the pandemic: conferencing tools, modes of remote communications (webcam/microphone), and locations of remote communications. Without priming participants, we surfaced the role of privacy and security in participants' decision making related to each aspect of remote communications.

\subsection{Conferencing Tools}
\begin{figure}[t]
\includegraphics[width=\columnwidth]{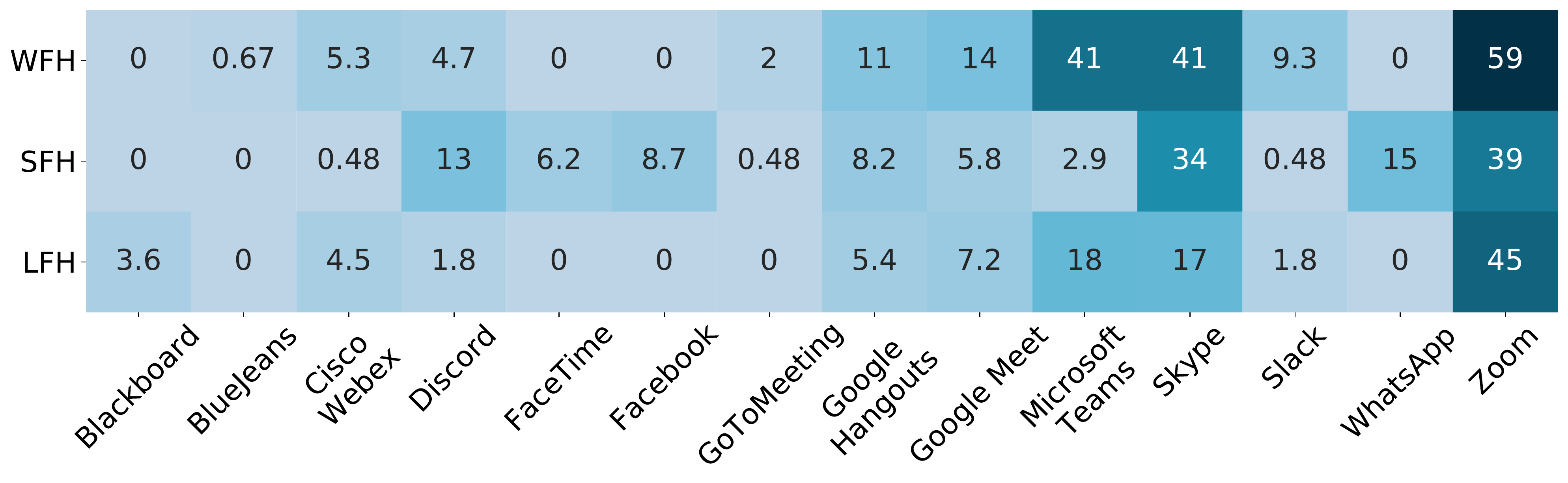}
    \caption{Usage of conferencing tools (in percentage) in different remote communication contexts reported by participants.}
    \label{fig:tool_heatmap}
\end{figure}

In all contexts, Zoom was reported to be used more frequently than other tools (WFH:~35\%, SFH:~27\%, and LFH:~42\%). We found that most participants (59\%) were using the same conferencing tool for their WFH and LFH meetings and 35\% of participants were using the same application in all three contexts. Figure~\ref{fig:tool_heatmap} shows the fraction of participants who reported using each of the conferencing tools at least once for their remote communications across the three contexts.

\begin{figure}[t]
\includegraphics[width=\columnwidth, trim = 0 0 0 0, clip]{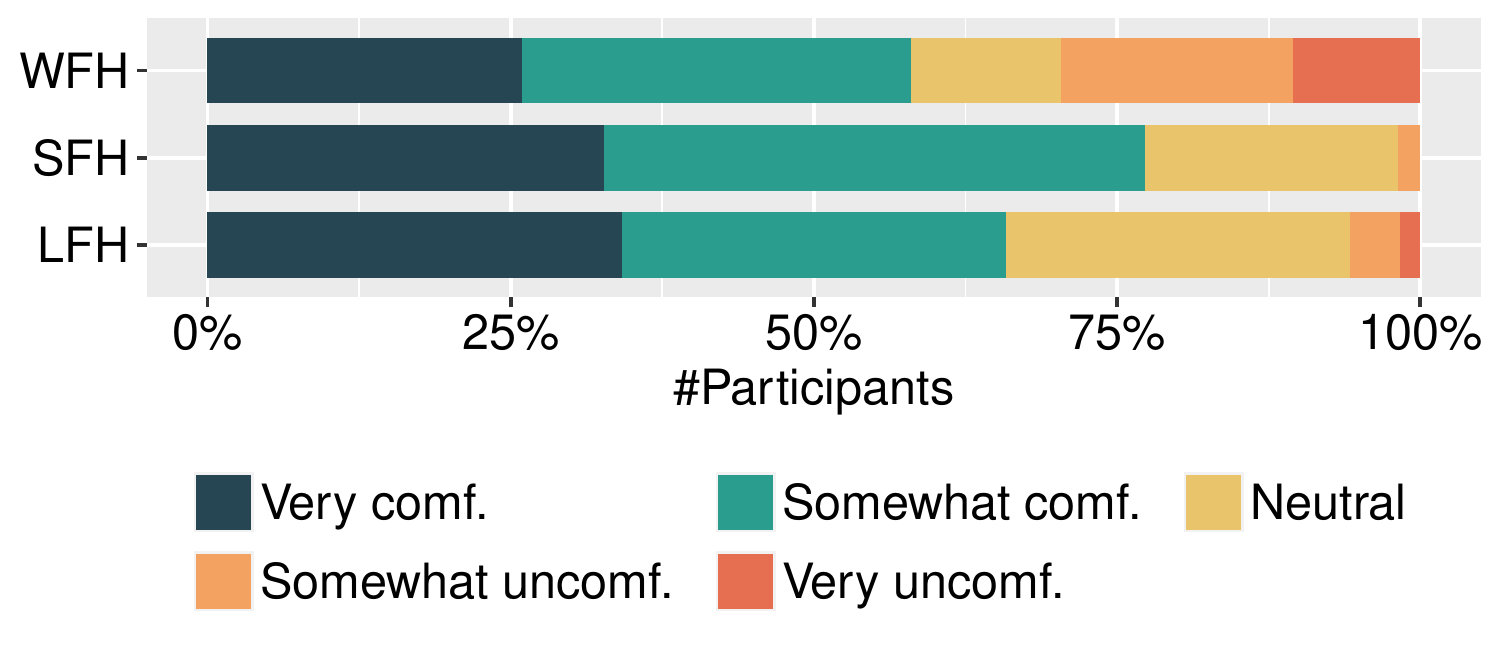}
    \caption{Participants' reported level of comfort with frequently used conferencing tools.}
    \label{fig:comfort_tool}
\end{figure}

In each context, on a five-point Likert scale, we asked participants to specify their level of comfort with the conferencing tool they most frequently use for their remote communications. Across all contexts, most participants (WFH:~87/150, SFH:~161/208, LFH:~75/114) were somewhat or very comfortable when using the conferencing tools for remote communications (see Figure~\ref{fig:comfort_tool}). Our regression analysis indicated that the context of remote communications significantly impacts participants' level of comfort (see Table~\ref{tab:regression}). We found that compared to work from home, participants were significantly more comfortable when using tools to communicate with family and friends (estimate $= 1.43$, $p$-value $< 0.05$) as well as communicating in the context of learning (estimate $= 0.88$, $p$-value $< 0.05$). Participants who reported to be using Google Hangouts were significantly more comfortable (estimate $= 2.13$, $p$-value $< 0.05$) with their conferencing tool than those who were using Zoom. In addition, our regression model suggested that an increase in the number of adults in the household resulted in a significant decline in the reported level of comfort with conferencing tools (estimate $= -1.03$, $p$-value $< 0.05$).

In each context, we explored participants' reasons behind their choice of conferencing tools as well as their comfort and discomfort with the tools. By qualitatively coding their open-ended responses, we surfaced several factors impacting participants' decision making and comfort level toward the use of conferencing tools.

\begin{table*}[t]
\centering\def\arraystretch{.95}
\setlength{\tabcolsep}{8.2pt}
\begin{tabular}{llllcccr}
\toprule
\textbf{\scriptsize Model No. and AIC} & \textbf{\scriptsize Dependent Variable} & \textbf{\scriptsize Independent Variable} & \textbf{\scriptsize Levels} & \textbf{\scriptsize Estimate} & \textbf{\scriptsize Odds Ratio} &\textbf{\scriptsize Std. Err.} & \textbf{\scriptsize $p$-value} \\ \hhline{--------}
\multirow{7}{*}{\normalfont \scriptsize 1 (AIC$=$271.58)} & \multirow{7}{*}{\normalfont \scriptsize Tool comfort level} & \multirow{2}{*}{\normalfont \scriptsize Context (baseline$=$WFH)} & \cellcolor{Gray}\normalfont \scriptsize SFH & \cellcolor{Gray}\normalfont \scriptsize 1.43 & \cellcolor{Gray}\normalfont \scriptsize 4.18 & \cellcolor{Gray}\normalfont \scriptsize 0.56 & {\cellcolor{Gray}\normalfont \scriptsize *} \\
& & & \cellcolor{Gray}\normalfont \scriptsize LFH & \cellcolor{Gray}\normalfont \scriptsize 0.88 & \cellcolor{Gray}\normalfont \scriptsize 2.41 & \cellcolor{Gray}\normalfont \scriptsize 0.44 & {\cellcolor{Gray}\normalfont \scriptsize *} \\ \cdashline{3-8}
& & \multirow{4}{*}{\normalfont \scriptsize Tool (baseline$=$Zoom)} & \cellcolor{Gray}\normalfont \scriptsize Google Hangouts & \cellcolor{Gray}\normalfont \scriptsize 2.13 & \cellcolor{Gray}\normalfont \scriptsize 8.41 & \cellcolor{Gray}\normalfont \scriptsize 0.81 & {\cellcolor{Gray}\normalfont \scriptsize *} \\
& & & \normalfont \scriptsize Google Meet & \normalfont \scriptsize 1.79 & \normalfont \scriptsize 5.99 & \normalfont \scriptsize 0.76 & \normalfont \scriptsize 0.30 \\
& & & \normalfont \scriptsize Microsoft Teams & \normalfont \scriptsize 0.98 & \normalfont \scriptsize 2.66 & \normalfont \scriptsize 0.76 & \normalfont \scriptsize 0.20 \\
& & & \normalfont \scriptsize Skype & \normalfont \scriptsize 0.80 & \normalfont \scriptsize 2.23 & \normalfont \scriptsize 0.74 & \normalfont \scriptsize 0.27 \\
\cdashline{3-8}
& & \normalfont \scriptsize \#Adults & \cellcolor{Gray}\normalfont \scriptsize $\{1,2,\dots\}$ & \cellcolor{Gray}\normalfont \scriptsize $-$1.03\phantom{$-$} & \cellcolor{Gray}\normalfont \scriptsize 0.36 & \cellcolor{Gray}\normalfont \scriptsize 0.52 & {\cellcolor{Gray}\normalfont \scriptsize *}\\
\hhline{--------}
\multirow{3}{*}{\normalfont \scriptsize 2 (AIC$=$266.71)} & \multirow{3}{*}{\normalfont \scriptsize Microphone usage} & \multirow{2}{*}{\normalfont \scriptsize Context (baseline$=$WFH)} & \cellcolor{Gray}\normalfont \scriptsize SFH & \cellcolor{Gray}\normalfont \scriptsize 1.53 & \cellcolor{Gray}\normalfont \scriptsize 4.61 & \cellcolor{Gray}\normalfont \scriptsize 0.39 & {\cellcolor{Gray}\normalfont \scriptsize ***} \\
& & & \cellcolor{Gray}\normalfont \scriptsize LFH & \cellcolor{Gray}\normalfont \scriptsize $-$1.61\phantom{$-$} & \cellcolor{Gray}\normalfont \scriptsize 0.20 & \cellcolor{Gray}\normalfont \scriptsize 0.33 & {\cellcolor{Gray}\normalfont \scriptsize ***} \\ \cdashline{3-8}
& & \multirow{1}{*}{\normalfont \scriptsize \#Children (7-13)} & \cellcolor{Gray}\normalfont \scriptsize $\{0,1,2,\dots\}$ & \cellcolor{Gray}\normalfont \scriptsize 1.29 & \cellcolor{Gray}\normalfont \scriptsize 3.63 & \cellcolor{Gray}\normalfont \scriptsize 0.44 & {\cellcolor{Gray}\normalfont \scriptsize **} \\ 
\hhline{--------}
\multirow{5}{*}{\normalfont \scriptsize 3 (AIC$=$322.47)} & \multirow{5}{*}{\normalfont \scriptsize Webcam usage} & \multirow{2}{*}{\normalfont \scriptsize Context (baseline$=$WFH)} & \cellcolor{Gray}\normalfont \scriptsize SFH & \cellcolor{Gray}\normalfont \scriptsize 1.66 & \cellcolor{Gray}\normalfont \scriptsize 5.26 & \cellcolor{Gray}\normalfont \scriptsize 0.34 & {\cellcolor{Gray}\normalfont \scriptsize ***} \\
& & & \cellcolor{Gray}\normalfont \scriptsize LFH & \cellcolor{Gray}\normalfont \scriptsize $-$1.49\phantom{$-$} & \cellcolor{Gray}\normalfont \scriptsize 0.22 & \cellcolor{Gray}\normalfont \scriptsize 0.34 & {\cellcolor{Gray}\normalfont \scriptsize ***} \\ \cdashline{3-8}
& & \multirow{2}{*}{\normalfont \scriptsize Age (baseline$=$18-29)} & \cellcolor{Gray}\normalfont \scriptsize 30-49 & \cellcolor{Gray}\normalfont \scriptsize 0.98 & \cellcolor{Gray}\normalfont \scriptsize 2.66 & \cellcolor{Gray}\normalfont \scriptsize 0.46 & {\cellcolor{Gray}\normalfont \scriptsize *} \\
& & & \normalfont \scriptsize 50-64 & \normalfont \scriptsize 2.71 & \normalfont \scriptsize 15.03 & \normalfont \scriptsize 2.05 & \normalfont \scriptsize 0.19 \\ 
\cdashline{3-8}
& & \multirow{1}{*}{\normalfont \scriptsize \#Rooms} & \cellcolor{Gray}\normalfont \scriptsize $\{0,1,2,\dots\}$ & \cellcolor{Gray}\normalfont \scriptsize 0.31  & \cellcolor{Gray}\normalfont \scriptsize 1.36 & \cellcolor{Gray}\normalfont \scriptsize 0.14 & {\cellcolor{Gray}\normalfont \scriptsize *} \\ 
\hhline{--------}
\multirow{8}{*}{\normalfont \scriptsize 4 (AIC$=$349.44)} & \multirow{8}{*}{\normalfont \scriptsize Location comfort level} & \multirow{2}{*}{\normalfont \scriptsize Context (baseline$=$WFH)} & \cellcolor{Gray}\normalfont \scriptsize SFH & \cellcolor{Gray}\normalfont \scriptsize 1.62 & \cellcolor{Gray}\normalfont \scriptsize 5.05 & \cellcolor{Gray}\normalfont \scriptsize 0.42 & {\cellcolor{Gray}\normalfont \scriptsize ***} \\
& & & \normalfont \scriptsize LFH & \normalfont \scriptsize 0.66 & \normalfont \scriptsize 1.93 & \normalfont \scriptsize 0.36 & \normalfont \scriptsize 0.06 \\ \cdashline{3-8}
& & \multirow{4}{*}{\normalfont \scriptsize Location (baseline$=$Bedroom)} & \normalfont \scriptsize Dining room & \normalfont \scriptsize $-$0.59\phantom{$-$} & \normalfont \scriptsize 0.55 & \normalfont \scriptsize 0.46 & \normalfont \scriptsize 0.44 \\
& & & \normalfont \scriptsize Kitchen & \normalfont \scriptsize $-$0.63\phantom{$-$} & \normalfont \scriptsize 0.53 & \normalfont \scriptsize 0.39 & \normalfont \scriptsize 0.52 \\
& & & \cellcolor{Gray}\normalfont \scriptsize Living room & \cellcolor{Gray}\normalfont \scriptsize $-$1.10\phantom{$-$} & \cellcolor{Gray}\normalfont \scriptsize 0.33 & \cellcolor{Gray}\normalfont \scriptsize 0.45 & {\cellcolor{Gray}\normalfont \scriptsize *} \\
& & & \normalfont \scriptsize Work room & \normalfont \scriptsize $-$0.38\phantom{$-$} & \normalfont \scriptsize 0.68 & \normalfont \scriptsize 0.54 & \normalfont \scriptsize 0.56 \\
\cdashline{3-8} 
& & \normalfont \scriptsize \#Adults & \cellcolor{Gray}\normalfont \scriptsize $\{1,2,\dots\}$ & \cellcolor{Gray}\normalfont \scriptsize $-$0.56\phantom{$-$} & \cellcolor{Gray}\normalfont \scriptsize 0.57 & \cellcolor{Gray}\normalfont \scriptsize 0.28 & {\cellcolor{Gray}\normalfont \scriptsize *}\\
\cdashline{3-8}
& & \normalfont \scriptsize Gender (baseline$=$Female) & \cellcolor{Gray}\normalfont \scriptsize Male & \cellcolor{Gray}\normalfont \scriptsize 1.04 & \cellcolor{Gray}\normalfont \scriptsize 2.83 & \cellcolor{Gray}\normalfont \scriptsize 0.42 & {\cellcolor{Gray}\normalfont ~\scriptsize *} \\
%\cdashline{1-8}
\cline{1-8}
\multicolumn{8}{c}{\normalfont \scriptsize \textit{Note:} \:\:\: *$p<$ 0.05 \:\:\: **$p<$ 0.01 \:\:\: ***$p<$ 0.001}\\
\bottomrule
\end{tabular}
\caption{Regression results of the CLMMs we built to explain participants' attitudes and concerns toward various remote communication aspects. A positive estimate of a level of an independent variable implies inclination toward an increase in the dependent variable and vice versa.}
\label{tab:regression}
\end{table*}

\subsubsection{Lack of Autonomy in Decision Making} \label{tool-autonomy}
In the contexts of WFH and LFH, participants frequently (WFH:~70/146, LFH:~64/107) implied that they have no agency over choosing what conferencing tool to use for their meetings. This lack of control was due to the fact that the tool was being selected for participants by their employers or educators, sometimes despite their personal preferences. 

In the WFH context, some participants (WFH:~21/70) reported that the required conferencing tool is aligned with privacy and security preferences and requirements of their employers. P16 reported to be using Microsoft Teams for their WFH meetings: \say{Work requires me to only use this tool. They say this is the most secure one out there.} Similarly P177 discussed why their employer asked them to use Microsoft Teams for WFH meetings: \say{It is the only tool that the company has approved security wise on our network.} 

In LFH, such imposed decisions contradicted some participants' (LFH:~17/64) personal privacy and security preferences, especially when they were required to use Zoom for their learning-related meetings. P36, who reported to use Zoom for their remote learning-related meetings, said: \say{That is the tool our teacher has chosen for us. Although security is certainly a problem.}

\subsubsection{Usability and Features} In all three contexts, the provided features and the usability of the tool were the second most frequently mentioned reasons to use the tool (WFH:~41/146, SFH:~92/204, LFH:~21/107) and the most commonly reported factors to make participants comfortable when using the conferencing tool (WFH:~75/111, SFH:~84/154, LFH:~42/70).  P9, who frequently uses Microsoft Teams for their work meetings, said: \say{I can clearly see every file that's been attached to our meetings, I can easily contact with others and the quality of voice and video is just perfect.} Unlike SFH, one of the most desirable features in the contexts of WFH and LFH was the ability of the tool to function properly with large groups. P102, who was using Zoom for their work meetings, said: \say{It supports a bigger number of people to be in a call better than the other ones.} 

In the WFH and LFH contexts, almost all participants (WFH:~40/41, LFH:~19/21) who mentioned using a conferencing tool based on its convenience and usability, reported to personally benefit the most from these attributes in their remote communications. P194 reported to be using Microsoft Teams for their WFH meetings: \say{Microsoft Teams allows me to stay connected more easily.}

Unlike WFH and LFH, in SFH, several participants reported to use a conferencing tool mainly due to its perceived ease of use and convenience for others on the call (e.g., family members or friends), especially those with limited familiarity with technology. P90, who was most frequently using Skype to communicate with family members, said: \say{Parents are not confident with tech, Skype was the easiest for them to set up.}

Some participants (WFH:~11/146, SFH:~9/204, LFH: 14/107) discussed giving up their privacy and security due to the tools' provided features and convenience. Almost all participants who mentioned such trade-offs reported to be using Zoom for their remote communications. In the context of WFH, P176 said: \say{Zoom offers the best features and is easy to use. Although security is certainly a problem.} Users' trade-off between privacy and security and provided convenience is a known behavior in the literature~\cite{zeng2019understanding, privacy-trade, emami2019exploring}.

\subsubsection{Familiarity with the Tool} The most mentioned reason in deciding what conferencing tool to use to communicate with friends and family members was how familiar the tool was to participants themselves and also others on the call (SFH:~94/204). P54 reported to use WhatsApp for their SFH meetings to accommodate their family members: \say{It's the one my family have already installed on their phones and know how to use.} Familiarity with the tool was the third most frequently mentioned reason in the contexts of WFH and LFH (WFH:~29/146, LFH:~15/107). P84 reported to use Skype more frequently than other tools: \say{The people I am calling with use Skype more than anything. I would rather use Zoom instead.} Familiarity was also the second most commonly mentioned contributor to participants' comfort with conferencing tools (WFH:~22/111, SFH:~48/154, LFH:~15/70). P12 reported why they are comfortable with using Skype for their SFH meetings: \say{I have been using Skype since I was a teenager, so I'm used to it and that makes me more comfortable when I'm talking to friends and family.}

Unlike WFH and LFH, in SFH 33\% of participants, who reported to value the familiarity with the tool the most, implied that such familiarity partially stemmed from using the tool in contexts other than socializing (e.g., work, learning). P10 reported to use Microsoft Teams for their SFH meetings: \say{My school uses the same platform and it's easier to be on only one platform at the same time.}

Participants' open-ended responses implied how familiarity with the tool impacted their privacy and security concerns. A few participants (WFH:~9/29, SFH:~17/94, LFH:~6/15) perceived a sense of safety when using the tool due to their prolonged experience with the tool. P30, who used Discord for their personal meetings, said: \say{I know it is very safe and reliable because I've been using it for the past 3 years.} This finding confirms the role of familiarity with technology in reducing risk perception~\cite{weber2005communicating}. Besides, a few participants associated their privacy concerns with their familiarity with the tool. P70 discussed why they only use Discord for their SFH meetings: \say{I already had account on it and also I am not comfortable sharing my info with more companies.}

\subsubsection{Privacy and Security Factors} \label{tool-privacy}
In all three contexts, the perceived privacy and security of the tool were the third most commonly mentioned factors in making participants comfortable when using the tool for their remote meetings (WFH:~22/111, SFH:~20/154, LFH:~11/70). In the context of WFH, Microsoft Teams and in SFH and LFH, Zoom were most frequently praised for their privacy and security practices.

When discussing their comfort with conferencing tools, some participants did not mention a specific privacy or security practice that made them comfortable when using the tool and instead said: \say{It is secure,} \say{It feels like a safe app,} or \say{I have no privacy concerns.} We qualitatively coded the open-ended responses and identified several privacy and security best practices and perceptions that were frequently reported across all contexts:

\begin{itemize}[leftmargin=*]
    \itemsep-0.3em 
    \item \textit{Information being encrypted (7)}: WhatsApp:3, Zoom:4
    \item \textit{Trusted brand (6)}: Microsoft Teams:2, Google Meet:1, Zoom:2, Cisco:1
    \item \textit{No reported risk on media (6)}: Microsoft Teams:3, Cisco:1, WhatsApp:1, Discord:1
    \item \textit{Protection from unauthorized access (5)}: Microsoft Teams:2, Google Meet:2, WhatsApp:1
    \item \textit{Ability to set password for meetings (2)}: Zoom:2
    \item \textit{No information being stored (1)}: Google Meet:1
\end{itemize}

Although most participants were comfortable with using the tools for their remote communications, some participants reported being somewhat or very uncomfortable (WFH:~17/150, SFH:~16/208, LFH:~7/114). Privacy and security were frequently mentioned as the reasons for participants' discomfort when using the tools (WFH:~8/17, SFH:~5/16, LFH: 3/7). Below is the list of privacy and security practices and beliefs that made participants uncomfortable when using the conferencing tool:

\begin{itemize}[leftmargin=*]
    \itemsep-0.3em 
    \item \textit{Risks and vulnerabilities reported by the media (7):} Zoom:7
    \item \textit{Personal space being exposed in the meeting (4):} Zoom:3, Google Meet:1
    \item \textit{Data being sold to third parties (2):} Zoom:2
    \item \textit{Amount of information being collected (2):} Google Duo:1, Skype:1
\end{itemize}

We asked participants to specify how they manage their reported discomfort with the conferencing tools. In the contexts of WFH and LFH, participants reported to address their concerns and discomfort by sharing less with other meeting attendees (WFH:~6/17, LFH:~2/7). Limiting the exposure was both in terms of restricting the content that is being shared, as well as modifying the configuration of their tool or the camera on their computer to limit the exposure. P4 limited the content they shared in the meeting and said: \say{I try not to say anything that could be used badly.} 

Unlike WFH and LFH, The most commonly mentioned mitigation approach in the context of SFH was limiting or avoiding the use of the tool (SFH:~9/16). P188, who reported using Google Hangouts, said: \say{I will uninstall it as soon as it is no longer needed.}

Some participants reported to take no action when being uncomfortable when using the tools (WFH:~4/17, SFH:~5/16, LFH:~1/7), mainly due to not being in charge of selecting the tools, not knowing what privacy and security controls the tool offers, or believing they have nothing to hide. P50 discussed why they do not take any action to address their privacy concerns with Google Duo: \say{Lots of other people use it too, nothing likely concerning will happen about what information the app ... collected on me.}

\subsection{Modes of Remote Communications} \label{AV-reason}
\begin{figure}[t]
\includegraphics[width=\columnwidth, trim = 0 0 0 0, clip]{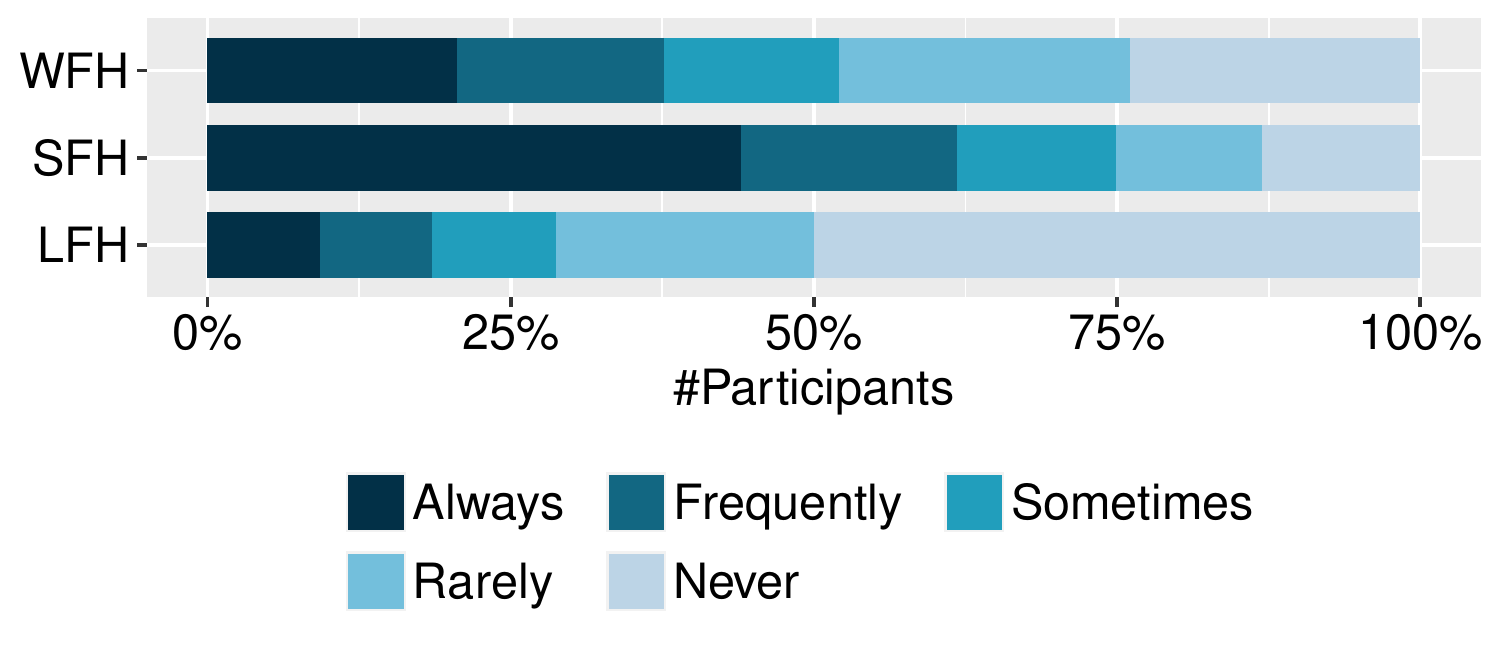}
    \caption{Reported frequency of using webcam.}
    \label{fig:webcam}
\end{figure}

\begin{figure}[t]
\includegraphics[width=\columnwidth, trim = 0 0 0 0, clip]{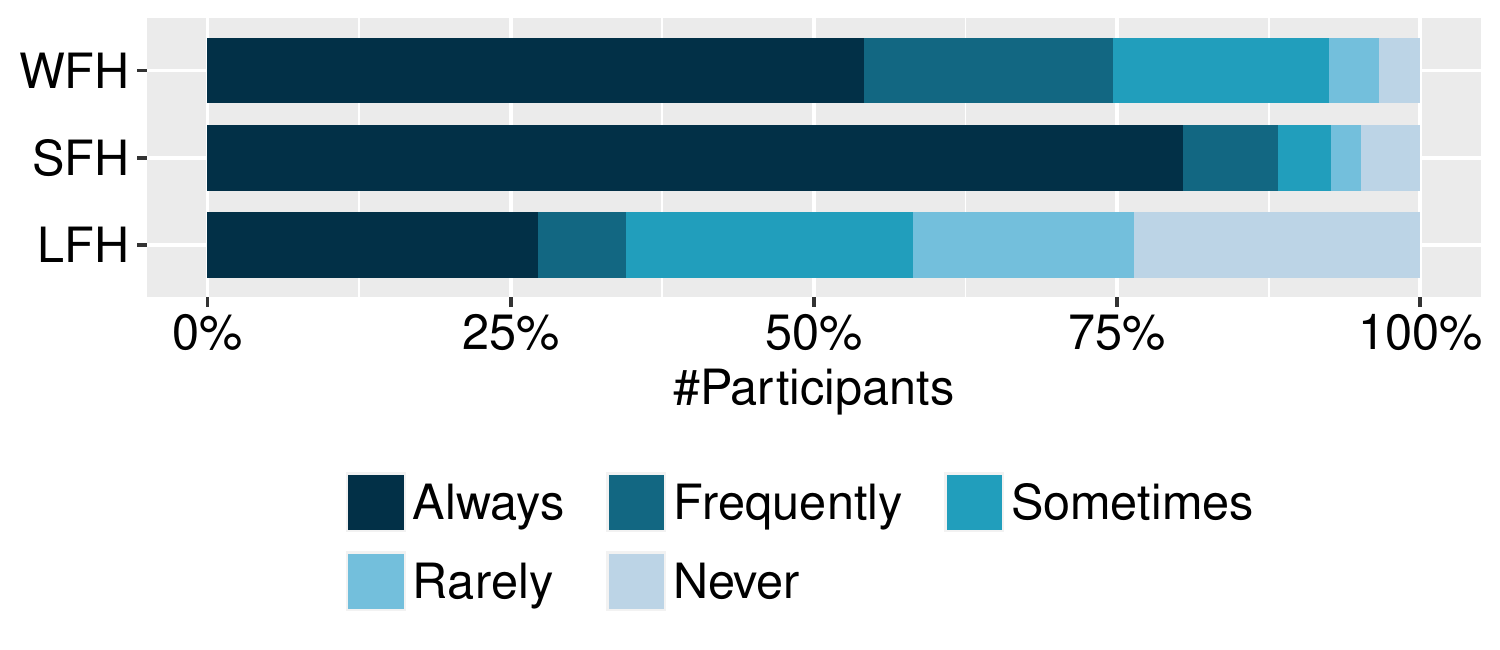}
    \caption{Reported frequency of using microphone.}
    \label{fig:microphone}
\end{figure}

We explored participants' attitudes and preferences toward the use of microphone and webcam in their remote communications. In all contexts, participants reported to activate their microphones significantly more frequently ($p$-value $< 0.001$) than their webcams when having remote communications (see Figures~\ref{fig:webcam} and~\ref{fig:microphone}). Our CLMM results showed that compared to WFH, participants turned on their webcams (estimate $= 1.66$, $p$-value $< 0.001$) and microphones (estimate $= 1.53$, $p$-value $< 0.001$) significantly more often when having remote personal meetings with friends and family members, and significantly less often (webcam usage: estimate $= -1.49$, $p$-value $< 0.001$; microphone usage: estimate $= -1.61$, $p$-value $< 0.001$) when having remote learning-related meetings. Other factors significantly impacting the frequency of webcam and microphone usage were the number of rooms in the home setting, age, and the number of children in the household (see Table~\ref{tab:regression}).

\subsubsection{Microphone/Webcam Misuse}
Accidental exposures of audio and video can lead to privacy violations. When asking participants about awkward incidents that had happened to them or others, across all contexts, the misuse of microphone and webcam was mentioned in almost all reported incidents (WFH:~29/34, SFH:~17/18, LFH:~8/10). In these incidents, the microphone and/or webcam were capturing unintended footage of a meeting attendee without their awareness and in some cases, without the awareness of the household members. For example, in the context of WFH, P194 mentioned an incident involving the microphone: \say{A member of management did not remember to mute himself while he answered his personal phone on speaker with what sounded to be a lawyer.} P185 reported a similar incident in the context of LFH that involved the misuse of webcam in the meeting: \say{Someone in a classroom stood up naked on the Zoom call and I guess he didn't know until it was too late.}

In order to raise users' risk awareness and prevent such incidents from happening, we need to understand the underlying reasons for participants' preferences towards different modes of remote communications. We asked participants how they decide to turn on their webcam and microphone when having remote communications.

\subsubsection{Agency over Decision Making} \label{mode-autonomy}
Participants' reasons to activate their microphone/webcam in WFH and LFH meetings implied their lack of agency over sharing their audio/video in their remote communications. In these contexts, respondents reported that they were explicitly expected to activate their microphone/webcam as a direct request by their employer or educator (WFH-Webcam: 73/101, LFH-Webcam: 66/94, WFH-Microphone: 77/127, LFH-Microphone: 59/97). Psychology literature refers to this type of behavior as \textit{obedience}, i.e., a form of social influence where group members change their behaviors and attitudes due to a direct request or command from an authority figure~\cite{cialdini2004social}. P101, who reported to always turn on the webcam in their remote work-related meetings, said: \say{There isn't a choice in terms of my manager requesting a meeting face to face.} For some participants, such imposed requests contradicted their personal preferences. P75 discussed their lack of desire to activate their webcam in work-related meetings: \say{If the manager asks me to turn on the video, I have to do it, but I personally prefer to maintain it switched off at all times.}

In the context of LFH, several participants reported being required to have their webcam and microphone on when taking exams (LFH-Webcam: 19/66, LFH-Microphone: 23/97). P43 discussed how they decide on when to activate their webcam during learning-related communications: \say{It depends if I'm explicitly asked by the professor to turn it on (For example when I have an \say{oral exam} since written exams are now hard to do remotely).}% Although these participants did not explicitly mention being asked to use remote proctoring tools, such tools have recently gained attention due to their invasive monitoring practices~\cite{proctor}.

Although no participant reported to be explicitly requested by others to activate their microphone/webcam in SFH meetings, our qualitative analysis found the implicit expectation to be the main factor in participants' microphone/webcam usage. Most participants perceived lack of control over the use of microphone/webcam and reported that they are naturally expected to turn on their microphone/webcam when talking to their family and friends (SFH-Webcam:~134/208, SFH-Microphone:~166/208). P183 reported: \say{I always turn the webcam on during personal meetings with friends and family because I think it is what they expect and it would be rude not to.}

Across all contexts, participants frequently mentioned that they would make the decision to activate their microphone/webcam based on other meeting attendees' behaviors (WFH-Webcam:~58/150, SFH-Webcam:~22/208, LFH-Webcam:~27/114, WFH-Microphone:~43/150, SFH-Microphone:~18/208, LFH-Microphone:~26/114). In the context of LFH, P185 reported to sometimes turn on their webcam: \say{If other students have their cameras on I am more likely to turn mine on. But if nobody has theirs on, I will probably not turn on my camera.} This type of social influence is called \textit{informational conformity}~\cite{deutsch1955study}, which serves as a cognitive repair~\cite{heath1998cognitive} and happens when group members follow others' behaviors and directions as they are unsure about the appropriate behavior~\cite{sherif1935study}. In several cases, participants reported to comply with the crowd despite holding a different preference. P43 reported to rarely turn on their webcam in work-related meetings: \say{Depending on the other person/people, and they always prefer to have a video. I personally find it a bit stressful, but don't mind it too much.}

In the context of SFH, some participants (SFH:~27/208) reported to jointly decide on the expectations around the use of microphone and webcam mostly prior to their personal meetings. P134 reported: \say{When we decide to meet, we choose video or none in the meeting invite.} A few participants discussed the importance of joint decision making in accommodating meeting attendees' preferences. P160 reported to frequently turn on their webcam when meeting their friends: \say{If I miss seeing her face, we will plan a video call. We plan them because she has anxiety which I definitely want to accommodate for as best [as I] can.}

\subsubsection{Attitudes over the Modes of Communication}
Our qualitative analysis indicated that participants shared diverse sentiments over activating their webcam/microphone in remote communications. In SFH, no participant suggested being uncomfortable with their lack of autonomy over webcam/microphone in their personal meetings. P117 discussed why they feel comfortable to always have their webcam on when meeting family and friends: \say{My family expect me to have video on, but I don't mind as I feel comfortable with people who really know me and accept me for who I am. I guess it's a gut feeling, if I don't feel anxious in their company in real life face to face, I would feel comfortable on a screen.} 

Unlike SFH, several participants (WFH:~21/101, LFH: 13/94) in WFH and LFH expressed negative attitudes toward having their webcam on. P80, who reported to activate their webcam at their employer's request, said: \say{I don't think video is necessary for the outcome of the meeting. It would be odd seeing colleagues in their home environment.} P84 discussed why they do not feel comfortable with having webcam on in their LFH meetings: \say{I do not want to be seen by people I have never met, so I do not turn it on, unless I am being asked by the teacher.} On the contrary, some participants (WFH:~16/101, LFH:~8/94) shared positive sentiments and supported having the webcam on during WFH and LFH meetings. P151, who reported to always turn on their webcam in WFH meetings, said: \say{I always turn it on as I feel face to face conversations with people create a better environment, and a higher level of honesty. I have campaigned for a policy in work to make video compulsory, and it has been taken up.} P43 discussed why they preferred to have their webcam on in LFH meetings: \say{It is \say{nice} and more productive during the Q\&A meetings to have webcam on and discuss about issues/doubts about a particular project.}

\subsection{Locations of Remote Communications}
\begin{figure}[t]
\includegraphics[width=\columnwidth]{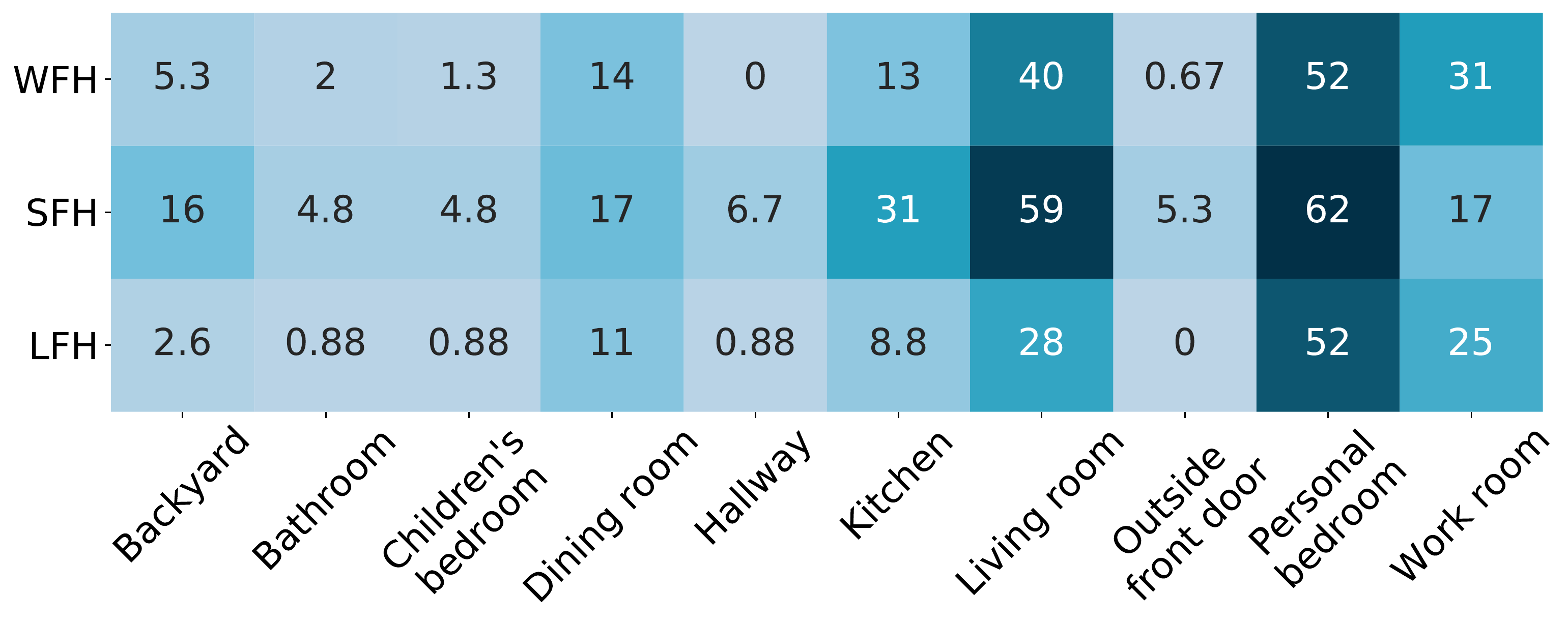}
    \caption{Usage of home locations (in percentage) reported by participants who engaged in remote communication contexts.}
    \label{fig:loc_heatmap}
\end{figure}

We asked participants to specify which part(s) of their homes they most frequently use for their remote meetings. Across the three contexts, participants' bedroom (WFH:~56/150, SFH:~83/208, LFH:~49/114), living room (WFH:~32/150, SFH:~75/208, LFH:~25/114), and study or workroom (WFH:~38/150, SFH:~19/208, LFH:~21/114) were reported to be used more often than other locations. Despite being rare, a few participants reported using their bathrooms for their remote meetings (WFH:~3/150, SFH:~10/208, LFH:~1/114). Figure~\ref{fig:loc_heatmap} shows the fraction of participants that reported to use each of the locations in their home at least once for their remote communications across the three contexts.

\begin{figure}[t]
\includegraphics[width=\columnwidth, trim = 0 0 0 0, clip]{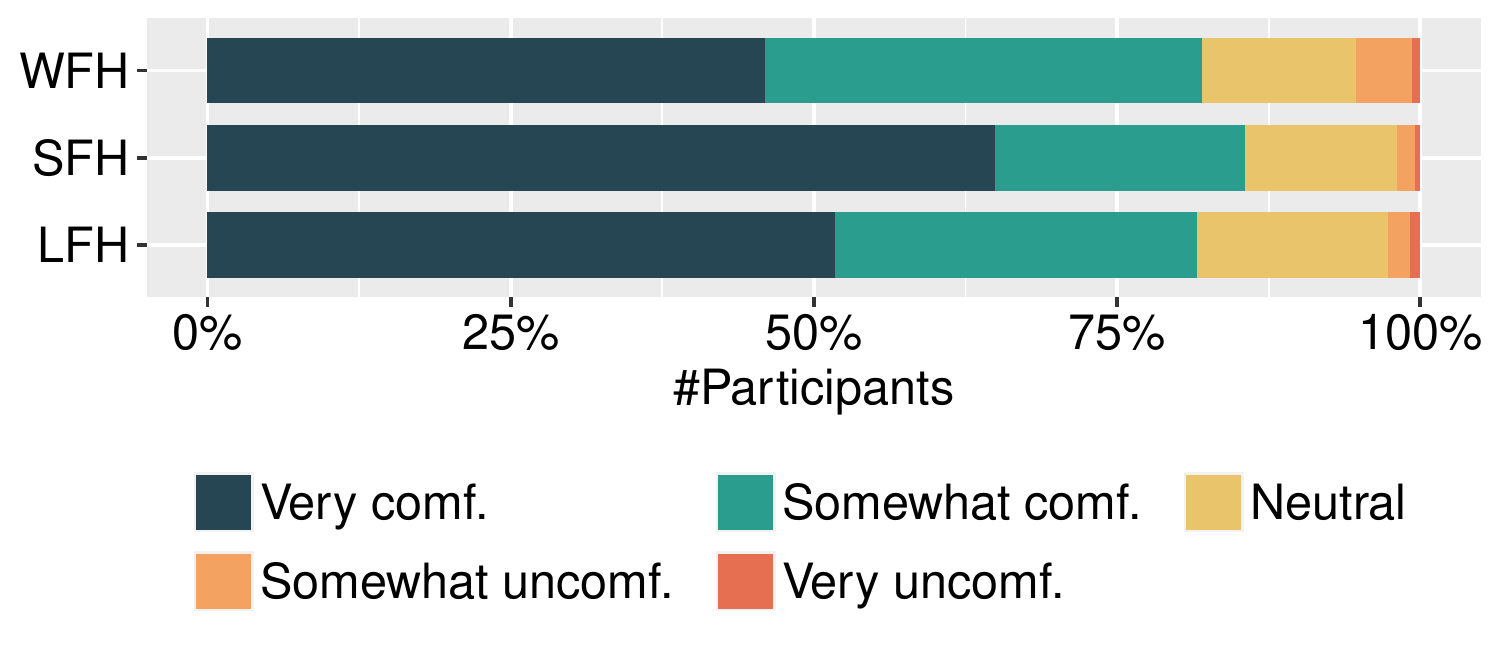}
    \caption{Participants' reported comfort level with frequently-used meeting locations.}
    \label{fig:comfort_loc}
\end{figure}

In all contexts, most participants (WFH:~123/150, SFH: 178/208, LFH:~91/114) were comfortable with the locations of their meetings (see Figure~\ref{fig:comfort_loc}) while having significant differences across the contexts. The regression analysis showed that compared to WFH, participants were significantly more comfortable when using any given location for their remote personal meetings (estimate $= 1.62$, $p$-value $< 0.001$). Other statistically significant factors to explain participants' level of comfort with their meeting locations were gender (estimate $= 1.04$, $p$-value $< 0.05$) and number of adults in the household (estimate $= -0.56$, $p$-value $< 0.05$). 

\subsubsection{Remote Privacy vs. Co-Inhabitant Privacy} \label{remote-co-privacy}
In all three contexts, several participants reported to select a meeting location which they perceived to be the most private in their homes. Especially, in the context of WFH, having privacy was the most frequently mentioned reason as to why a location is used for work meetings (WFH:~42/147). In the contexts of LFH and SFH, privacy was the second and third most common reason for participants' location-related decision making, respectively (LFH:~24/114, SFH:~35/208). Moreover, we found privacy and sense of safety to be frequently reported (WFH:~42/123, SFH:~68/178, LFH:~24/91) as participants' reasons to be comfortable with their meeting locations. 

By qualitatively analyzing participants' open-ended responses, we identified two types of privacy: remote privacy and co-inhabitant privacy. Remote privacy refers to having privacy from other meeting attendees, while co-inhabitant privacy refers to having privacy from other household members. The types of privacy that were mentioned by participants varied among different contexts of remote communications.

In the context of SFH, participants reported to have no concern over remote privacy as they felt comfortable with other meeting attendees (e.g., friends, family members) viewing their personal space. For example, P32 reported why they feel comfortable holding their personal meetings in the living room: \say{It's my living room, it's organized and everyone I talk to already knows it.} P196, who reported to use their bedroom for remote personal meetings, said: \say{I feel very comfortable in my bedroom as friends generally know my place.}

Almost all participants who considered privacy when selecting their SFH meeting locations reported to be concerned with their co-inhabitant privacy. These participants reported to choose their meeting locations to be a personal space in their home where they are not being disturbed or interrupted by others during their remote meetings. Participants' personal space was mostly reported to be their bedroom. A few participants reported to have co-inhabitant privacy in other locations, including living room, kitchen, and workroom. P185 reported to use their bedroom for SFH meetings: \say{I feel the most comfortable in my own bedroom and I know that it will be a private space and that the least amount of interruptions will happen in my bedroom compared to other areas of the house.} P25 reported why they were using the living room for personal meetings: \say{It is separated from the room in which my housemate works and so less likely that he will overhear.} 

Unlike SFH, in the contexts of WFH and LFH, participants' perception of privacy was more diverse. Some participants (WFH:~19/42, LFH:~15/24) reported to prefer having co-inhabitant privacy by detaching their meeting locations from other household members. P47 discussed why they decided to use their bedroom for remote work-related meetings: \say{I prefer using mostly my bedroom for most of my online work/meetings as I don't like others hearing me talk, I like to have a little bit of \say{privacy}.} P43 reported using their bedroom to preserve their co-inhabitant privacy: \say{I like to have my own little space, a bit of privacy from the rest of family and less distractions around so I can focus on the course.} On the other hand, some participants were uncomfortable about others on the call seeing their personal space. These participants reported to desire having remote privacy by selecting a less personal location that provides a \say{neutral} or \say{professional} background with fewer details about their personal life. P195 talked about having privacy when holding their work meetings in the study or workroom: \say{I don't have anything private there that I would be unprofessional if I had to share my webcam with others.} P122 discussed why they were using the kitchen for remote learning meetings: \say{This is the least personal place in the house to have such meetings.}

Open-ended responses revealed potential conflicts between remote privacy and co-inhabitant privacy. In all contexts, participants who valued their co-inhabitant privacy frequently reported to be using their personal bedroom to have privacy from their household members. At the same time, participants in WFH and LFH perceived their personal bedrooms to be intimate and, therefore, not appropriate to preserve their remote privacy. We found similar tensions with regard to using living room for remote communications. Some participants chose to hold their remote meetings in the living room to have remote privacy, although having less co-inhabitant privacy due to the interruptions by household members. 

\subsubsection{Room Convenience and Equipment} Similar to participants' attitudes toward conferencing tools, convenience and comfort were frequently mentioned as the deciding factor when selecting a meeting location (WFH:~26/147, SFH:~65/208, LFH:~24/114). Especially, in the context of SFH, participants reported that the convenience of the location is the most important reason when choosing the room for their remote personal meetings. P80 discussed why they use the living room for their personal meetings: \say{This is the location of relaxation and the area where my husband and I can sit comfortably and talk to friends and family.}

Another commonly mentioned reason behind participants' choice of meeting location in all three contexts was the presence of equipment that was needed for remote communications, including computer, desk, and books (WFH:~42/147, SFH:~41/208, LFH:~37/114). For participants in the context of LFH, the room equipment was the main factor in deciding what room to use for their learning meetings. P75, who reported to use their bedroom for LFH meetings, said: \say{This is the location where I have my desk and my PC in.}

\subsubsection{Discomfort with Meeting Locations} 
Although most participants were comfortable with their selected meeting locations, some respondents reported to be somewhat or very uncomfortable (WFH:~9/150, SFH:~4/208, LFH:~3/114). Across all contexts, the main factor participants mentioned that made them uncomfortable with a location was the perceived invasion of remote/co-inhabitant privacy when holding meetings there (WFH:~4/9, SFH:~1/4, LFH:~2/3). In the LFH context, P115 reported that they are uncomfortable with using their bedroom for remote learning-related meetings: \say{It is hard to get comfortable in the bedroom as it feels like a private area to invite people in to.} P4 discussed their discomfort with using the living room for their remote work meetings: \say{I could be overheard and am not comfortable with the webcam being on as it intrudes on my privacy.}

We asked participants how they manage their perceived discomfort with meeting locations. The open-ended responses indicated that only participants in the WFH context took steps to mitigate their discomfort, while in the contexts of SFH and LFH, participants reported to take no action when being uncomfortable with their remote meeting locations. The primary approach participants mentioned to take in the context of WFH was to limit the information exposure, either to other meeting attendees or their household members (WFH:~3/9). P38, who reported to mainly use their living room for their work-related meetings, said: \say{I minimise what can be seen and test the audio quality before the meeting.} Similarly, P35 limited the work-related information from the household members: \say{I close my door and ask other family members [not to] come to the living room when having work meetings.}
\section{Discussion} \label{sec:discussion}
We first provide a brief comparison between the contexts of remote communications. Based on our findings, we then discuss methods to inform and enable users' privacy-protective decision making related to remote communications. 

\subsection{Context-Specific Privacy Concerns and Attributes}
We focused on three remote communication contexts: working from home (WFH), socializing from home (SFH), and learning from home (LFH). In each context, we surfaced participants' attitudes and concerns toward the use of remote communication technologies. Our quantitative and qualitative findings suggested several similarities and differences in participants' attitudes, behaviors, and privacy concerns among the three contexts. In all contexts, comfort and discomfort with conferencing tools and meeting locations were mainly explained by participants' privacy and security concerns and their perceived sense of safety. Our findings indicated that WFH and LFH were similar in terms of the choice and the use of conferencing tools (e.g., activating webcam/microphone). In SFH, unlike other contexts, the decisions toward the conferencing tools and the meeting locations were primarily based on the provided convenience.

Numerous articles have been published that provide recommendations on how to better protect privacy when engaging in remote communications~\cite{hygiene, overnight, protect1}. Almost all of these guidelines are targeted toward the users, who are already struggling with an insurmountable mental pressure thanks to the pandemic. When an awkward incident happens in a conference call, end users are not the only group to blame, as they are only a small part of the remote communication ecosystem. Tool developers and users' employers and educators could play a critical role in informing and empowering users to adopt privacy-protective behaviors while communicating with others online.

The pandemic may not last forever, but remote communications will stay longer~\cite{after-survey} and that requires us to critically examine what we have learned during the pandemic. Based on our findings, in the following, we distill several recommendations to inform and empower users, and to design more privacy-protective tools. 

\subsection{Enabling Context-Specific Informed Decision Making}
Participants' open-ended responses showed lack of autonomy in their attitudes and behaviors toward remote communication technologies. Several participants reported to have no control over the choice of conferencing tools for their WFH and LFH meetings (see Section~\ref{tool-autonomy}). Lack of active decision making was also apparent in participants' attitudes toward the use of webcam and microphone. Participants reported to be explicitly (WFH and LFH) or implicitly (SFH) expected to turn on/off their webcam/camera in the meetings (see Section~\ref{mode-autonomy}). The qualitative findings indicated that having limited or no control over the conferencing tools and their features (e.g., webcam/microphone) was participants' primary impediment to managing their tool-related privacy and security concerns (see Section~\ref{tool-privacy}). To enable active and informed decision making in remote communications, we need to consider the context of the meeting. 

Our findings suggested that WFH and LFH meetings have similar power dynamics that are being set by an authority figure (e.g., employer, educator). By providing \textbf{inclusive, transparent, and flexible policies}, workplaces and education institutes can take the first step toward informing and empowering meeting participants. To be inclusive, policies should acknowledge users' diverse and context-specific privacy needs and attitudes. To provide holistic privacy-protective policies, future studies should be conducted to explore other stakeholders' perspectives of remote communications, including but not limited to, employers and teachers.

In light of our findings, organizational policies need to discuss the choice of conferencing tool, the use of microphone and webcam in the meetings, and the available user controls. In addition, the policies should be flexible and open for feedback to help meeting attendees discuss and manage their concerns and discomfort. Items to be outlined in such policies include:

\begin{itemize}[leftmargin=*]
    \itemsep-0.3em 
    \item What conferencing tools should be used for the meetings and why? 
    \item What privacy and security controls are provided by the tools?
    \item In what condition are users (not) required to use their microphone/webcam?
    \item How can users control their microphone/webcam in the communication tools?
    \item How can meeting participants manage their concerns and discomfort with the tools?
\end{itemize}

Compared to WFH and LFH, in the context of SFH, participants felt being more in control of choosing a conferencing tool, which might be partially due to more balanced power dynamics. However, because of the implicit expectations, several participants felt having no control over the decision to activate their webcam/microphone in the personal meetings. As recommended by a few of our participants, \textbf{joint decision making} prior to the meeting could give meeting participants the opportunity to discuss their concerns and decide on a policy that accommodates and respects all of them. 

\subsection{Inclusive Privacy by Design}
Across all contexts, the main factor participants mentioned to make them uncomfortable with a meeting location was the lack of remote and co-inhabitant privacy they felt when holding remote meetings in that location (see Section~\ref{remote-co-privacy}). Participants who referred to remote privacy reported that they do not feel comfortable having their home locations in the background of their WFH and LFH meetings. On the other hand, having a neutral or generic background was one of the frequently mentioned factors to make participants comfortable when using a meeting location (see Section~\ref{remote-co-privacy}).

Due to the restrictions posed by the diverse working, living, and learning arrangements, it may not be reasonable to ask everyone to find a neutral background for their remote meetings. Tool developers can enable features to help users protect their privacy. Some of the current communication tools, such as Zoom~\cite{zoom-background} and Microsoft Teams~\cite{teams-background}, already allow users to cover their real background by using virtual ones. Similarly, tools such as Skype~\cite{skype-blur} and Google Meet~\cite{google-blur} provide a feature for users to blur their backgrounds.

Across all contexts, when discussing co-inhabitant privacy, several participants reported to be uncomfortable with other household members hearing their conversations (see Section~\ref{remote-co-privacy}). From the regression analysis, we found that an increase in the number of household members leads to a significant decrease in the level of comfort with conferencing tools as well as the locations of remote communications (see Table~\ref{tab:regression}). To protect people's privacy in different contexts of remote communications, we need to design for diverse household settings. For example, to preserve co-inhabitant privacy in crowded settings, future remote communication devices can be enabled with a feature to detect and notify the user whether other household members are in the hearing range of their remote meetings. Such features can also respect the privacy needs of other meeting attendees, e.g., in case meeting participants are not comfortable with their voice or video being heard or seen by individuals who are not part of the call (e.g., household members).

\subsection{Limitations}
As the first paper to study remote communications at the transition of the pandemic, we surfaced participants' attitudes, behaviors, and concerns toward specific aspects of remote communications in different contexts. Due to the focus of our research and the survey methodology, we did not explore other potentially informative research questions, which could be studied in the future. In what follows, we will highlight the limitations of the current work, alongside several future research directions.

Our study used Prolific to recruit survey participants. Prior work recommended using Prolific to recruit a diverse sample of participants~\cite{peer2017beyond}. However, despite its diverse population and similarly to other crowdsourcing platforms, Prolific participants are not representative of any average population. For example, in Prolific, participants tend to be younger and more educated~\cite{prolific-limit}. In addition, our participants were mainly from the UK, Poland, and the US, and we had a small number of participants from other countries (see Table~\ref{tab:demographics}). Due to these limitations, the findings of our study should not be generalized. Our study provides an overview of technology-related perceptions and behaviors during the global COVID-19 pandemic and we believe future studies can more directly focus on specific populations. In our study, participants' country of residence was not a statistically significant factor, which might be due to the small number of participants from some of the countries. Future studies could explore the difference in privacy concerns and attitudes among different countries and cultures.

As we previously mentioned, among other questions, our survey explored how participants' learning experience has been impacted by the COVID-19 pandemic. In this study, we only recruited participants who were at least 18 years old. However, it is also important to understand the impact of the pandemic and the privacy considerations of students from all ages, which should be considered in a future study.

To ensure participants' familiarity with the contexts of remote communications, for each context, we only asked the survey questions of participants who reported to have familiarity with that specific context. This potential selection bias might impact participants' attitudes and concerns toward remote communications in each context. Similarly, due to the nature of the job and depending on the level of experience, crowd-source participants might be more familiar with remote communication technologies than the average population. Having familiarity with a technology has been shown to decrease the amount of risk an individual would perceive related to that technology~\cite{weber2005communicating}. Therefore, the reported privacy and security concerns captured by our study could be lower than the average population' risk perception toward remote communications.
\section{Conclusion}\label{sec:conclusion}
The COVID-19 pandemic has caused people around the world to abruptly shift their in-person work, personal life, and/or education meetings to remote ones, which could outlast the pandemic. Therefore, to enable safe remote experience, it is critical to design privacy-protective tools and empower users to consider privacy and security when engaging in remote communications. To this end, we conducted a 220-participant survey on Prolific, in which we considered three contexts of remote communications, namely working (WFH), socializing (SFH), and learning from home (LFH). Our quantitative and qualitative findings indicated that concerns, attitudes, and behaviors toward remote communications are diverse and context-dependent. Across all contexts, privacy and security were among the most frequently mentioned concerns that participants had. These concerns were exacerbated by the fact that participants felt that they had no agency over decision making about conferencing tools as well as audio and video modes of remote communications. We provided several recommendations for tool developers and organizations to enable users to make privacy- and security-protective choices when engaging with remote communications.

\section*{Acknowledgments}
We are especially grateful to our study participants. We thank Christine Geeng for reading the draft of our paper. We are also very thankful to our reviewers for their valuable feedback. This work was supported in part by the NSF awards CNS-1565252 and CNS-1651230 and the University of Washington Tech Policy Lab, which receives support from the William and Flora Hewlett Foundation, the John D. and Catherine T. MacArthur Foundation, Microsoft, and the Pierre and Pamela Omidyar Fund at the Silicon Valley Community Foundation.

{
\interlinepenalty=10000

\Urlmuskip=0mu plus 1mu\relax
\def\UrlBreaks{\do\/\do_\do-}

\bibliographystyle{IEEEtran}
\bibliography{bibliography}
}

\appendix
\section{Survey Questions} \label{survey_questions}
\subsection{Informed Consent} \label{consent}
This is a survey about technology use in the home during the Coronavirus (COVID-19) pandemic by researchers at the University of Washington, in Seattle, Washington, USA. The University of Washington’s Human Subjects Division reviewed our study, and determined that it was exempt from federal human subjects regulation. We do not expect that this survey will put you at any risk for harm.

In order to participate, you must be at least 18 years old and able to complete the survey in English. We expect this survey will take about 20 minutes to complete. If you have any questions about this survey, you may email us at \linebreak \href{mailto:hometechnology@cs.washington.edu}{hometechnology@cs.washington.edu}.

\begin{itemize}
    \item I am 18 years or older. \\ $\circ$ Yes \: $\circ$ No
    \item I have read and understand the information above. \\ $\circ$ Yes \: $\circ$ No
    \item I want to participate in this research and continue with the task. \\ $\circ$ Yes \: $\circ$ No
\end{itemize}

\subsection{Context-Specific Questions (CQ)} \label{CQ}
In the contexts of WFH, SFH, and LFH, we referred to remote communications as \say{remote work-related meetings,} \say{remote personal meetings with friends and family members,} and \say{remote learning-related meetings,} respectively. Here we only provide the questions for the WFH context.

\begin{itemize}
    \item \textbf{CQ1:} Have you been mostly having remote work-related meetings from home during the COVID-19 pandemic? \\ $\circ$ Yes \: $\circ$ No 
    
    \textit{The rest of the context-related questions will only be presented if the answer is “Yes.”}
    
    \item \textbf{CQ2:} Before the COVID-19 pandemic, how often have you had remote work-related meetings from home? \\ $\circ$ Never \: $\circ$ Once or twice a year \: $\circ$ Once every 4-6 months \: $\circ$ Once every 2-3 months \: $\circ$ Once every month \: $\circ$ Once every 2-3 weeks \: $\circ$ Once every week \: $\circ$ Not every day, but more than once a week \: $\circ$ Every day
    
    \item \textbf{CQ3:} During the COVID-19 pandemic, how many hours a week do you spend in remote work-related meetings from home? \\ $\circ$ Less than 1 \: $\circ$ 1 to 5 hours \: $\circ$ 6 to 10 hours \: $\circ$ 11 to 15 hours \: $\circ$ 16 to 20 hours \: $\circ$ 21 to 25 hours \: $\circ$ 26 to 30 hours \: $\circ$ 31 to 35 hours \: $\circ$ 36 to 40 hours \: $\circ$ Over 40 hours
    
    \item \textbf{CQ4:} During the COVID-19 pandemic, how long have you been having remote work-related meetings from home? \\ $\circ$ Since last week \: $\circ$ Since two weeks ago \: $\circ$ Since three weeks ago \: $\circ$ Since one month ago \: $\circ$ Since more than one month ago
    
    \item \textbf{CQ5:} During the COVID-19 pandemic, what conferencing tools do you mostly use for your remote work-related meetings? If you use more than one tool, please select the one you use most frequently. \\ $\circ$ BlueJeans \: $\circ$ Google Hangouts \: $\circ$ Google Meet \: $\circ$ GoToMeeting \: $\circ$ Microsoft Teams \: $\circ$ Skype \: $\circ$ Slack \: $\circ$ UberConference \: $\circ$ Zoom \: $\circ$ Other (please specify [Open-ended])
    
    \item \textbf{CQ6:} Please explain why you have been using the tool that you have specified more frequently than other tools. [Open-ended]

    \item \textbf{CQ7:} During the COVID-19 pandemic, in your current environment, how do you feel about using this tool for your remote work-related meetings? \\ $\circ$ Very uncomfortable \: $\circ$ Somewhat uncomfortable \: $\circ$ Neither uncomfortable nor comfortable \: $\circ$ Somewhat comfortable \: $\circ$ Very comfortable

    \item \textbf{CQ8:} \textit{(If in CQ7, Very uncomfortable or Somewhat uncomfortable is selected)} What about this tool makes you uncomfortable when using it? [Open-ended]

    \item \textbf{CQ9:} \textit{If in CQ7, Very uncomfortable or Somewhat uncomfortable is selected)} How do you manage your discomfort when using this tool? [Open-ended]

    \item \textbf{CQ10:} \textit{If in CQ7, Very comfortable or Somewhat comfortable is selected)} What about this tool makes you comfortable when using it? [Open-ended]

    \item \textbf{CQ11:} During the COVID-19 pandemic, how often do you turn on your device's webcam when having remote work-related meetings? \\ $\circ$ Never \: $\circ$ Rarely \: $\circ$ Sometimes \: $\circ$ Frequently \: $\circ$ Always
    
    \item \textbf{CQ12:} How do you decide whether or not to turn on your device's webcam when having remote work-related meetings? [Open-ended]

    \item \textbf{CQ13:} During the COVID-19 pandemic, how often do you turn on your device's microphone when having remote work-related meetings? \\ $\circ$ Never \: $\circ$ Rarely \: $\circ$ Sometimes \: $\circ$ Frequently \: $\circ$ Always
    
    \item \textbf{CQ14:} How do you decide whether or not to turn on your device's microphone when having remote work-related meetings? [Open-ended]

    \item \textbf{CQ15:} During the COVID-19 pandemic, which area of your home do you usually hold your remote work-related meetings in? If you use more than one location, please select the one you use most frequently for remote work-related remote meetings. \\ $\circ$ Backyard \: $\circ$ Bathroom \: $\circ$ Bedroom (yours) \: $\circ$ Bedroom (your children's) \: $\circ$ Dining room \: $\circ$ Hallway \: $\circ$ Kitchen \: $\circ$ Living room \: $\circ$ Outside front door \: $\circ$ Study or workroom \: $\circ$ Other (please specify [Open-ended])

    \item \textbf{CQ16:} During the COVID-19 pandemic, how do you feel about using this location to have remote work-related meetings? \\ $\circ$ Very uncomfortable \: $\circ$ Somewhat uncomfortable \: $\circ$ Neither uncomfortable nor comfortable \: $\circ$ Somewhat comfortable \: $\circ$ Very comfortable
    
    \item \textbf{CQ17:} \textit{(If in CQ16, Very uncomfortable or Somewhat uncomfortable is selected)} What about this location makes you uncomfortable when having remote work-related meetings there? [Open-ended]

    \item \textbf{CQ18:} \textit{If in CQ16, Very uncomfortable or Somewhat uncomfortable is selected)} How do you manage your discomfort when using this location for having remote work-related meetings? [Open-ended]

    \item \textbf{CQ19:} \textit{If in CQ16, Very comfortable or Somewhat comfortable is selected)} What about this location makes you comfortable when having remote work-related meetings there? [Open-ended]
    
    \item \textbf{CQ20:} During the COVID-19 pandemic, have you or people you know ever experienced an awkward incident while having remote work-related meetings? \\ $\circ$ Yes \: $\circ$ No 

    \item \textbf{CQ21:} \textit{(If in CQ18, Yes is selected)} Please describe the incident. [Open-ended]
    \item \textbf{CQ22:} \textit{(If in CQ18, Yes is selected)} Please describe what you or people you know have done in response to the incident. [Open-ended]
\end{itemize}

\subsection{IoT-Related Questions (IQ)} \label{IQ}
\begin{itemize}
    \item \textbf{IQ1:} How many smart home devices (e.g., smart speakers, smart security cameras) do you currently have at home? If you have more than one unit for each type of device, please count them all. \\ $\circ$ 0 \: $\circ$ 1 \: $\circ$ 2 \: $\circ$ 3 \: $\circ$ 4 \: $\circ$ More than 4
    
    \item \textbf{IQ2:} \textit{(If in IQ1, 0 is not selected)} What smart home devices do you currently have? [Open-ended]
    
    \item \textbf{IQ3:} \textit{(If in IQ1, 0 is not selected)} In which rooms of your home are each of your smart devices located? Please specify the location of all the smart devices you currently have. [Open-ended]

    \item \textbf{IQ4:} \textit{(If in IQ1, 0 is not selected)} Have you made any changes regarding your smart devices during the COVID-19 pandemic (e.g., changes to behaviors around devices, changes of device locations, and changes to device settings)? \\ $\circ$ Yes \: $\circ$ No 

    \item \textbf{IQ5:} \textit{(If in IQ4, Yes is selected)} Please explain the changes you have made. [Open-ended]

    \item \textbf{IQ6:} \textit{(If in IQ1, 0 is not selected)} During the COVID-19 pandemic, my privacy-related concerns about the data collection by my smart home devices \\ $\circ$ have strongly increased compared to before the pandemic \: $\circ$ have slightly increased compared to before the pandemic \: $\circ$ have not changed compared to before the pandemic \: $\circ$ have slightly decreased compared to before the pandemic \: $\circ$ have strongly decreased compared to before the pandemic

    \item \textbf{IQ7:} \textit{(If in IQ6, Have strongly increased [...] or Have slightly increased [...] is selected)} Please explain why your privacy-related concerns about the data collection by your smart home devices have increased compared to before the pandemic. [Open-ended]
    
    \item \textbf{IQ8:} \textit{(If in IQ6, Have strongly decreased [...] or Have slightly decreased [...] is selected)} Please explain why your privacy-related concerns about the data collection by your smart home devices have decreased compared to before the pandemic. [Open-ended]
    
    \item \textbf{IQ9:} \textit{(If in IQ6, Have not changed [...] is selected)} Please explain why your privacy-related concerns about the data collection by your smart home devices have not changed compared to before the pandemic. [Open-ended]
\end{itemize}

\subsection{Social Media Related Questions (SMQ)} \label{SMQ}
\begin{itemize}
    \item \textbf{SMQ1:} How many social media platforms (e.g., Facebook, Twitter) are you a member of? \\ $\circ$ 0 \: $\circ$ 1 \: $\circ$ 2 \: $\circ$ 3 \: $\circ$ 4 \: $\circ$ More than 4
    
    \item \textbf{SMQ2:} \textit{(If in SMQ1, 0 is not selected)} What social media platforms are you a member of? [Open-ended]
    
    \item \textbf{SMQ3:} \textit{(If in SMQ1, 0 is not selected)} Have you made any changes about how you use the social media platforms during the COVID-19 pandemic? \\ $\circ$ Yes \: $\circ$ No 
    
    \item \textbf{SMQ4:} \textit{(If in SMQ3, Yes is selected)} Please explain the changes you have made. [Open-ended]
    
    \item \textbf{SMQ5:} \textit{(If in SMQ1, 0 is not selected)} During the COVID-19 pandemic, my privacy-related concerns about using social media platforms \\ $\circ$ have strongly increased compared to before the pandemic \: $\circ$ have slightly increased compared to before the pandemic \: $\circ$ have not changed compared to before the pandemic \: $\circ$ have slightly decreased compared to before the pandemic \: $\circ$ have strongly decreased compared to before the pandemic

    \item \textbf{SMQ6:} \textit{(If in SMQ5, Have strongly increased [...] or Have slightly increased [...] is selected)} Please explain why your privacy-related concerns about using social media platforms have increased compared to before the pandemic. [Open-ended]
    
    \item \textbf{SMQ7:} \textit{(If in SMQ5, Have strongly decreased [...] or Have slightly decreased [...] is selected)} Please explain why your privacy-related concerns about using social media platforms have decreased compared to before the pandemic. [Open-ended]
    
    \item \textbf{SMQ8:} \textit{(If in SMQ5, Have not changed [...] is selected)} Please explain why your privacy-related concerns about using social media platforms have not changed compared to before the pandemic. [Open-ended]
\end{itemize}

\subsection{Demographics and Home Settings} \label{DH}
\begin{itemize}
    \item \textbf{DH1:} Including yourself, how many adults 18 years of age and above live in your current home? \\ $\circ$ 1 \: $\circ$ 2 \: $\circ$ 3 \: $\circ$ 4 \: $\circ$ 5 \: $\circ$ More than 5
    
    \item \textbf{DH2:} How many children at or above the age of 13 and under the age of 18 live in your current home? \\ $\circ$ 0 \: $\circ$ 1 \: $\circ$ 2 \: $\circ$ 3 \: $\circ$ 4  \: $\circ$ 5 \: $\circ$ More than 5
    
    \item \textbf{DH3:} How many children at or above the age of 7 and under the age of 13 live in your current home? \\ $\circ$ 0 \: $\circ$ 1 \: $\circ$ 2 \: $\circ$ 3 \: $\circ$ 4  \: $\circ$ 5 \: $\circ$ More than 5
    
    \item \textbf{DH4:} How many children under the age of 7 live in your current home? \\ $\circ$ 0 \: $\circ$ 1 \: $\circ$ 2 \: $\circ$ 3 \: $\circ$ 4  \: $\circ$ 5 \: $\circ$ More than 5
    
    \item \textbf{DH5:} Who do you share your home with? (check as many as apply) \\ $\circ$ No one \: $\circ$ Roommate(s) \: $\circ$ Spouse(s)/Domestic partner(s) \: $\circ$ Children \: $\circ$ Parent(s) \: $\circ$ Other (please specify [Open-ended])

    \item \textbf{DH6:} Do you have shared wall(s) with your neighbors? \\ $\circ$ Yes \: $\circ$ No
    
    \item \textbf{DH7:} How many bedrooms does your home have? \\ $\circ$ 0 \: $\circ$ 1 \: $\circ$ 2 \: $\circ$ 3 \: $\circ$ 4  \: $\circ$ 5 \: $\circ$ More than 5
    
    \item \textbf{DH8:} How many rooms other than bedrooms does your home have? \\ $\circ$ 0 \: $\circ$ 1 \: $\circ$ 2 \: $\circ$ 3 \: $\circ$ 4  \: $\circ$ 5 \: $\circ$ More than 5
    
    \item \textbf{DH9:} What is your age? \\ $\circ$ 18-29 years old \: $\circ$ 30-49 years old \: $\circ$ 50-64 years old \: $\circ$ 65 years and older

    \item \textbf{DH10:} What is your gender? [Open-ended]
    
    \item \textbf{DH11:} What is the highest degree you have earned? \\ $\circ$ No schooling completed \: $\circ$ Nursery school \: $\circ$ Grades 1 through 11 \: $\circ$ 12th grade—no diploma \: $\circ$ Regular high school diploma \: $\circ$ GED or alternative credential \: $\circ$ Some college credit, but less than 1 year of college \: $\circ$ 1 or more years of college credit, no degree \: $\circ$ Associates degree (for example: AA, AS) \: $\circ$ Bachelor’s degree (for example: BA. BS) \: $\circ$ Master’s degree (for example: MA, MS, MEng, MEd, MSW, MBA) \: $\circ$ Professional degree beyond bachelor’s degree (for example: MD, DDS, DVM, LLB, JD) \: $\circ$ Doctorate degree (for example: Ph.D., EdD)

    \item \textbf{DH12:} In which country do you currently reside? [List of countries provided by Qualtrics]
    
    \item \textbf{DH13:} What is your current employment status? \\ $\circ$ Full-time employment \: $\circ$ Part-time employment \: $\circ$ Unemployed \: $\circ$ Self-employed \: $\circ$ Home-maker \: $\circ$ Student \: $\circ$ Retired
    
    \item \textbf{DH14:} \textit{(If in DH13, Unemployed or Retired is not selected)} The organization you work for is in which of the following? \\ $\circ$ Public sector (e.g., government) \: $\circ$ Private sector (e.g., most businesses and individuals) \: $\circ$ Non-for-profit sector

    \item \textbf{DH15:} Do you have a background in technology? \\ $\circ$ Yes \: $\circ$ No
    
    \item \textbf{DH16:} \textit{(If in DH14, Yes is selected)} Please specify what your technical background is. [Open-ended]
\end{itemize}

\section{Smart Home IoT Devices} \label{smart-home}
Privacy and security concerns toward smart home IoT devices are not unique to the pandemic. However, their associated privacy risks have increased as people tend to spend more time at home~\cite{smart-hack}. In our study, we explored the impact of the pandemic on participants' attitudes and privacy concerns toward their smart home devices.

\subsection{Background}
As people spend more time at home due to the pandemic, the demand for smart home IoT devices has seen a sharp surge. In a recent survey conducted by Parks Associates, 33\% of smart home device owners reported an increase in the usage of their smart home devices during the pandemic compared to before the pandemic~\cite{surge-usage-IoT}. In another study, more than half of the participants reported purchasing new smart home devices during the pandemic~\cite{surge-purchase-IoT}.

The risks and vulnerabilities of smart home devices are not new to privacy and security researchers. Prior to the COVID-19 pandemic, several articles have reported on an array of privacy and security challenges and issues related to smart home devices~\cite{sivaraman2015network, alexa_out, google, alexa, foscam, bugeja2016privacy}. During the pandemic, due to the surge in purchasing and using smart home devices, many of which having poor privacy and security practices~\cite{unsafe-market}, users are now exposing themselves and potentially others to an even higher level of risk. Reports have shown that attacks against IoT devices have more than doubled as compared to before the pandemic~\cite{smart-hack}. IoT consumers increasingly set up their smart home devices on the same network as their work-related computers, consequently increasing their workplaces' attack surface~\cite{smart-network}. Moreover, a study has shown that smart speakers can be mistakenly triggered up to 19 times per day and on each activation, keep a recording of up to 43 seconds~\cite{dubois2020speakers}, which is clearly disconcerting when having sensitive work, social, or learning-related meetings.

Reports published prior to the pandemic showed that consumers of smart devices are concerned about the privacy and security practices of such devices~\cite{granjal2015security, arias2015privacy, sicari2015security, moz}. As an example, a survey by Consumers International and the Internet Society reported that 63\% of respondents found the data collection of smart devices to be \say{creepy} and more than half did not trust their devices to respect their privacy~\cite{creepy-iot}. 

Literature has discussed consumers' IoT-related privacy and security concerns prior to the pandemic and no study has been conducted to explore the impact of the pandemic on users' privacy and security concerns and attitudes. As part of our survey, we asked questions to understand how the pandemic has changed participants' privacy perception and behavior related to their smart home devices.

\subsection{Results}

Out of 103 participants who reported to have at least one smart home device, 74\% had a smart speaker with voice assistant (e.g., Amazon Echo, Google Home). Other common smart home devices that participants had were smart security camera (23/103), smart TV (22/103), and smart light bulb (12/103). Most participants reported to place their smart devices in their living room (57/103), about half placed them in the bedroom (49/103), some in the kitchen (31/103), and a few in the bathroom (16/103). 

\subsubsection{Attitudes and Concerns} \label{smart-concern}
Without prompting participants about the privacy and security of smart home devices, we asked them whether they changed anything about their smart home devices during the pandemic. Only four participants reported to have made changes to their smart home devices and almost all reported to become more aware of the presence of the microphones on smart speakers and to turn them off when they are not needed. P174 reported: \say{During the pandemic, I've been more diligent about turning the microphone off, especially in my work/study room, so it doesn't get triggered by errant phrases or record sensitive conversations.} Similarly, P160 discussed the change they applied to their Google Home Mini that is placed in the their bedroom: \say{I've left voice recognition off. I really don't like that anyone can ask it to do things, I don't want to be shocked awake in the morning.}

To capture the impact of the pandemic on privacy-related concerns, we asked participants to specify how their concerns toward the data collection of their smart home devices have been impacted by the pandemic. In addition, we asked participants to provide the rationale behind their answers. Most participants (78/103) reported that the pandemic has not impacted their IoT-related privacy concerns, due to several reasons.

\boldpartitle{Lack of Risk Awareness}
Participants' justifications as to why the pandemic has not impacted their privacy concerns suggested their lack of risk awareness as the most frequent reason (38/78). Participants commonly reported that they have no reason to change their concerns mainly due to not having heard about or personally experienced new risks during the pandemic. P123 discussed why their privacy concerns have not been impacted: \say{I haven't seen any disturbing news about privacy during a pandemic.} Some participants attributed their lack of awareness to having the same knowledge about the risks of their smart home devices compared to before the pandemic. P39 reported: \say{I knew the potential risks before and I still do. I haven't seen anything new that would change my opinion.}

\boldpartitle{Lack of Initial Concern}
Several participants reported that they always had little or no privacy concerns with the data practices of their smart home devices and that the pandemic has not impacted that (16/78). Most of these participants reported to have at least one smart speaker with voice assistant (11/16) and some had smart TVs (5/16). P34, who had a smart TV in their living room, said: \say{I have never been concerned about it so the pandemic has not changed it. I believe that mass data is necessary and helpful overall.}

Although the pandemic did not impact most participants' privacy concerns toward their smart home IoT devices, some participants (15/78) reported that their IoT-related privacy concerns have grown during the pandemic, primarily due to increased awareness and interaction with their devices.

\boldpartitle{Increased Risk Awareness}
Participants frequently reported that they became more aware of the privacy risks of their smart home devices during the pandemic (11/15). Some of these participants attributed their heightened risk awareness to the increased privacy-focused media reports during the pandemic (6/11). P146 reported: \say{I have heard so much about privacy issues on the news because of the quarantine so now I'm a little more worried about how much data companies track.}

\boldpartitle{More Frequent Device Usage}
The second reason contributing to participants' increased privacy concerns was spending more time around the smart home devices and having more interactions with them during the pandemic (8/15). This increase in device usage is aligned with a recent survey conducted by Parks Associates~\cite{surge-usage-IoT}. P175, who had a smart speaker and smart TV in their living room, reported: \say{We use them more during the pandemic, and although they are not devices that can collect a lot of personal info, the worries are still there.} P174, who had a smart speaker in their workroom, discussed their increased privacy concerns: \say{I don't want it to be recording sensitive work information that I discuss aloud.}

\section{Social Media Platforms} \label{social-media}
During the pandemic, social media platforms have seen an unprecedented growth in user engagement~\cite{social}. This increase in interaction could increase users' exposure to online risks, including phishing attacks, identity theft, and cyberbulling~\cite{saravanakumar2016privacy, chewae2015much, privacy-social}. We explored the impact of the pandemic on participants' attitudes, privacy concerns, and risk awareness toward their social media platforms. 

\subsection{Background}
Similar to smart home IoT devices, the use of social media platforms has been soaring due the COVID-19 pandemic. Stay at home orders have caused more people using social media platforms to stay connected with their colleagues, family, and friends. A study showed that during the pandemic, social media platforms have seen a 61\% increase in usage~\cite{social}. Such an increase has led privacy experts and lawmakers to be concerned about users' privacy during the pandemic~\cite{tik-tok-risk, tik-tok-ban}.

In addition to an increase in usage, the pandemic might have potentially exposed social media users to a greater harm by impacting the content people share due to their change in risk assessment when posting online~\cite{nabity2020inside}. The change in usage and content sharing on social media exacerbates the already alarming privacy risks and concerns of these platforms, including identity theft attacks, phishing attacks, cyberbulling, and malware attacks~\cite{saravanakumar2016privacy, chewae2015much, privacy-social}. However, from the literature, it is unclear whether users acknowledge the intensified risks and harms of social media during the pandemic and how their risk perception has impacted their behaviors toward such platforms. Our study contributes to the literature by investigating how the pandemic has influenced users' concerns, attitudes, and behaviors toward their social media platforms.

\subsection{Results}

Almost all (214/220) of our participants reported to be active on at least one social media platform. The most frequently used platforms were Facebook (174/214), Instagram (151/214), Twitter (119/214), and Reddit (86/214).

\subsubsection{Attitudes and Concerns} \label{media-concern}
19\% of participants (40/214) who were active on at least one social media platform reported to change how they use their social media platforms during the pandemic. Most participants reported to have increased their interactions with social media (28/40). P56, who was active on Facebook, said: \say{I have been on it more, accessing news. I also use Facebook's messenger feature to chat with friends and ex-colleagues.} Some participants, on the other hand, reported to lessen their interactions with social media. P141, who reported to have accounts on several platforms, said: \say{I use social media less since people were panicking about COVID-19 in the beginning and I didn't want to consume so much stressful content.}

We asked participants to specify how the pandemic has impacted their privacy-related concerns toward social media platforms. Most participants reported that the pandemic did not impact their privacy concerns (155/204). Quite surprisingly, 68\% of participants (18/28) who increased their interactions with social media platforms during the pandemic, reported that their privacy concerns with social media has not been impacted. Open-ended responses suggested several reasons as to why privacy concerns toward social medial platforms were not impacted during the pandemic. 

\boldpartitle{Lack of Risk Awareness}
Lack of risk knowledge and awareness was the most frequently reported reason as to why some participants' privacy concerns toward social media platforms have not been impacted during the pandemic (61/155). P71, who reported to be active on Facebook, Twitter, and Instagram said: \say{I am not aware of social media platforms changing their data practices over the pandemic. Thus, my concern stays the same on those platforms.} P39 attributed their lack of awareness to media reports: \say{I haven't seen any news recently that would change my knowledge about the potential risks of my social media accounts.}

\boldpartitle{Perceived Efficacy}
To some participants (27/155), their unaffected privacy concerns were mainly due to their perception of having autonomy over protecting their privacy. The two common approaches participants reported to adopt to manage their privacy on social media were posting less sensitive information and changing the privacy settings (e.g., who can view the posts). P86 discussed how they protect their privacy by limiting their online activities: \say{I knew from the start what were the privacy-related risks of social media. That's why I don't really post sensitive information anymore.} P176 talked about their privacy-protective social media settings: \say{I already have my privacy settings tightly controlled so have had no need to adjust them.}

\boldpartitle{Lack of Initial Concern} Similar to IoT-related privacy concerns, some participants (25/155) reported that their privacy concerns related to their social media platforms have not been changed during the pandemic as they have never had any concern about such platforms. P24, who had Twitter and Facebook accounts, said: \say{I have no concerns about my accounts so they haven't changed during the pandemic.}

\boldpartitle{Unchanged User Behavior}
Some respondents (25/155) reported that the pandemic has not affected their concerns toward the privacy practices of their social media platforms since their interactions with their accounts stayed the same as compared to before the pandemic. P95, who reported to have LinkedIn and Facebook accounts, said: \say{I use them in the same way as before and I didn't see any difference so no need for extra concern.}

A few participants (25/204) reported that they became more concerned about their social media platforms during the pandemic, mainly due to heightened risk awareness and increased platform usage.

\boldpartitle{Increased Risk Awareness} Several participants reported that during the pandemic, they became more aware of the privacy risks of social media platforms (19/25). P23 discussed their privacy concerns related to oversharing on social media during the pandemic: \say{I am now more aware that when you are stressed or sad you tend to overshare, especially if you do not have any other outlets, and I am concerned for people especially during the pandemic time since I have seen that many do not realize they share a lot of sensitive information that can be easily collected by other people.} P61 mentioned how they became more aware of the potential privacy harms of social media during the pandemic: \say{I have had more time to think how my privacy can be breached through social media, so I have deleted information and closed my privacy settings more.}

\boldpartitle{Increased Platform Usage} Increased interactions with social media platforms reported to be the second most frequent reason as to why some participants became more concerned with their online platforms (14/25). P175, who had Reddit, Discord, and Twitter accounts, said: \say{I'm using social media a lot more, and people have tried to get into my e-mail, and I have personal information on some accounts.} P4, who reported being active on Facebook and Instagram, discussed why they became more concerned with social media: \say{These days more people are using social media, so they might become easier to hack.}

\section{Codebooks} \label{codebooks}
{\urlstyle{sf}
The codebooks are available at:\\ {\color{note_blue}\url{https://gist.github.com/SOUPS-COVID-Privacy/97b6f6caeb13d5091314e6458049617d}}.
}

\vfill\eject

\section{Participants' Information}\label{demographics}
\begin{table}[h]
\setlength{\tabcolsep}{5pt}
\centering\def\arraystretch{.95}
\begin{tabular}{llrrr}
\toprule
\multirow{2}{*}{\footnotesize Timeline} &
\multirow{2}{*}{\footnotesize Meeting Frequency} & \multicolumn{3}{c}{\footnotesize Context} \\ \cmidrule(lr){3-5}
& & \footnotesize WFH & \footnotesize SFH & \footnotesize LFH \\ \hline
\multirow{9}{*}{\textnormal{\footnotesize Before the pandemic}} & \textnormal{\footnotesize Never} & \textnormal{\footnotesize 53\%} & \textnormal{\footnotesize 33\%} & \textnormal{\footnotesize 60\%} \\
& \textnormal{\footnotesize Once/twice a year} & \textnormal{\footnotesize 6\%} & \textnormal{\footnotesize 7\%} & \textnormal{\footnotesize 14\%} \\
& \textnormal{\footnotesize Once every 4-6 months} & \textnormal{\footnotesize 4\%} & \textnormal{\footnotesize 6\%} & \textnormal{\footnotesize 5\%} \\
& \textnormal{\footnotesize Once every 2-3 months} & \textnormal{\footnotesize 4\%} & \textnormal{\footnotesize 7\%} & \textnormal{\footnotesize 3\%} \\
& \textnormal{\footnotesize Once every month} & \textnormal{\footnotesize 3\%} & \textnormal{\footnotesize 6\%} & \textnormal{\footnotesize 2\%} \\
& \textnormal{\footnotesize Once every 2-3 weeks} & \textnormal{\footnotesize 4\%} & \textnormal{\footnotesize 10\%} & \textnormal{\footnotesize 2\%} \\
& \textnormal{\footnotesize Once every week} & \textnormal{\footnotesize 6\%} & \textnormal{\footnotesize 13\%} & \textnormal{\footnotesize 3\%} \\
& \textnormal{\footnotesize > once a week} & \textnormal{\footnotesize 11\%} & \textnormal{\footnotesize 12\%} & \textnormal{\footnotesize 5\%} \\
& \textnormal{\footnotesize Every day} & \textnormal{\footnotesize 9\%} & \textnormal{\footnotesize 6\%} & \textnormal{\footnotesize 6\%} \\ \hdashline[3pt/2pt] 
\multirow{10}{*}{\textnormal{\footnotesize During the pandemic}} & \textnormal{\footnotesize < 1 (hour/week)} & \textnormal{\footnotesize 25\%} & \textnormal{\footnotesize 35\%} & \textnormal{\footnotesize 11\%} \\
& \textnormal{\footnotesize 1-5 (hour/week)} & \textnormal{\footnotesize 44\%} & \textnormal{\footnotesize 41\%} & \textnormal{\footnotesize 40\%} \\
& \textnormal{\footnotesize 6-10 (hour/week)} & \textnormal{\footnotesize 13\%} & \textnormal{\footnotesize 14\%} & \textnormal{\footnotesize 16\%} \\
& \textnormal{\footnotesize 11-15 (hour/week)} & \textnormal{\footnotesize 7\%} & \textnormal{\footnotesize 3\%} & \textnormal{\footnotesize 12\%} \\
& \textnormal{\footnotesize 16-20 (hour/week)} & \textnormal{\footnotesize 6\%} & \textnormal{\footnotesize 2\%} & \textnormal{\footnotesize 11\%} \\
& \textnormal{\footnotesize 21-25 (hour/week)} & \textnormal{\footnotesize 1\%} & \textnormal{\footnotesize 1\%} & \textnormal{\footnotesize 7\%} \\
& \textnormal{\footnotesize 26-30 (hour/week)} & \textnormal{\footnotesize 1\%} & \textnormal{\footnotesize 1\%} & \textnormal{\footnotesize 3\%} \\
& \textnormal{\footnotesize 31-35 (hour/week)} & \textnormal{\footnotesize 2\%} & \textnormal{\footnotesize 1\%} & \textnormal{\footnotesize 0\%} \\
& \textnormal{\footnotesize 36-40 (hour/week)} & \textnormal{\footnotesize 1\%} & \textnormal{\footnotesize 1\%} & \textnormal{\footnotesize 0\%} \\
& \textnormal{\footnotesize > 40 (hour/week)} & \textnormal{\footnotesize 0\%} & \textnormal{\footnotesize 1\%} & \textnormal{\footnotesize 0\%} \\
\bottomrule
\end{tabular}
\caption{Frequency of engaging in remote communications.}
\label{tab:meeting freq.}
\end{table}

% \vspace{.5in}

\begin{table*}[h]
\setlength{\tabcolsep}{12pt}
\centering
\begin{tabular}{lrrr}
\toprule
\multirow{2}{*}{\footnotesize Experience Duration} & \multicolumn{3}{c}{\footnotesize Context} \\ \cmidrule(lr){2-4}
& \footnotesize WFH & \footnotesize SFH & \footnotesize LFH \\ \hline
\textnormal{\footnotesize Since last week} & \textnormal{\footnotesize 1\%} & \textnormal{\footnotesize 1\%} & \textnormal{\footnotesize 7\%} \\
\textnormal{\footnotesize Since two weeks ago} & \textnormal{\footnotesize 0\%} & \textnormal{\footnotesize 1\%} & \textnormal{\footnotesize 5\%} \\
\textnormal{\footnotesize Since three weeks ago} & \textnormal{\footnotesize 3\%} & \textnormal{\footnotesize 2\%} & \textnormal{\footnotesize 9\%} \\
\textnormal{\footnotesize Since one month ago} & \textnormal{\footnotesize 5\%} & \textnormal{\footnotesize 3\%} & \textnormal{\footnotesize 18\%} \\
\textnormal{\footnotesize Since more than one month ago} & \textnormal{\footnotesize 91\%} & \textnormal{\footnotesize 93\%} & \textnormal{\footnotesize 61\%} \\
\bottomrule
\end{tabular}
\caption{Summary statistics of how long participants were experiencing the three contexts under study.}
\label{tab:since}
\end{table*}

\begin{table*}[h]
\setlength{\tabcolsep}{6pt}
\centering
\begin{tabular}{lrrrrrrrr}
\toprule
\footnotesize Question & \multicolumn{8}{c}{\footnotesize Responses} \\ \hline
\textnormal{\footnotesize Shared wall(s) with} & \textnormal{\footnotesize Yes} & \textnormal{\footnotesize No} & & & & & \\
\textnormal{\footnotesize neighbors} & \textnormal{\footnotesize 54\%} & \textnormal{\footnotesize 46\%} & & & & & & \\ \hdashline[3pt/2pt]
\multirow{2}{*}{\textnormal{\footnotesize Housemates}} & \textnormal{\footnotesize No one} & \textnormal{\footnotesize Roommate(s)} & \textnormal{\footnotesize Spouse(s)/Domestic partner(s)} & \textnormal{\footnotesize Children} & \textnormal{\footnotesize Parent(s)} & \textnormal{\footnotesize Other: Siblings} \\
& \textnormal{\footnotesize 3\%} & \textnormal{\footnotesize 3\%} & \textnormal{\footnotesize 19\%} & \textnormal{\footnotesize 11\%} & \textnormal{\footnotesize 21\%} & \textnormal{\footnotesize 25\%} \\ \hdashline[3pt/2pt]
\multirow{2}{*}{\textnormal{\footnotesize \#Adults 18+ years old}} & \textnormal{\footnotesize 1} & \textnormal{\footnotesize 2} & \textnormal{\footnotesize 3} & \textnormal{\footnotesize 4} & \textnormal{\footnotesize 5} & \textnormal{\footnotesize More than 5} \\
& \textnormal{\footnotesize 12\%} & \textnormal{\footnotesize 45\%} & \textnormal{\footnotesize 22\%} & \textnormal{\footnotesize 17\%} & \textnormal{\footnotesize 4\%} & \textnormal{\footnotesize 0\%} \\ \hdashline[3pt/2pt]
\textnormal{\footnotesize \#Children between} & \textnormal{\footnotesize 0} & \textnormal{\footnotesize 1} & \textnormal{\footnotesize 2} & \textnormal{\footnotesize 3} & \textnormal{\footnotesize 4} & \textnormal{\footnotesize 5} & \textnormal{\footnotesize More than 5} \\
\textnormal{\footnotesize 13 and 18 years old}& \textnormal{\footnotesize 81\%} & \textnormal{\footnotesize 12\%} & \textnormal{\footnotesize 6\%} & \textnormal{\footnotesize 1\%} & \textnormal{\footnotesize 0\%} & \textnormal{\footnotesize 0\%} & \textnormal{\footnotesize 0\%} \\ \hdashline[3pt/2pt]
\textnormal{\footnotesize \#Children between} & \textnormal{\footnotesize 0} & \textnormal{\footnotesize 1} & \textnormal{\footnotesize 2} & \textnormal{\footnotesize 3} & \textnormal{\footnotesize 4} & \textnormal{\footnotesize 5} & \textnormal{\footnotesize More than 5} \\
\textnormal{\footnotesize 7 and 13 years old}& \textnormal{\footnotesize 86\%} & \textnormal{\footnotesize 14\%} & \textnormal{\footnotesize 0\%} & \textnormal{\footnotesize 0\%} & \textnormal{\footnotesize 0\%} & \textnormal{\footnotesize 0\%} & \textnormal{\footnotesize X\%} \\ \hdashline[3pt/2pt]
\textnormal{\footnotesize \#Children under} & \textnormal{\footnotesize 0} & \textnormal{\footnotesize 1} & \textnormal{\footnotesize 2} & \textnormal{\footnotesize 3} & \textnormal{\footnotesize 4} & \textnormal{\footnotesize 5} & \textnormal{\footnotesize More than 5} \\
\textnormal{\footnotesize 7 years old}& \textnormal{\footnotesize 83\%} & \textnormal{\footnotesize 12\%} & \textnormal{\footnotesize 5\%} & \textnormal{\footnotesize 0\%} & \textnormal{\footnotesize 0\%} & \textnormal{\footnotesize 0\%} & \textnormal{\footnotesize 0\%} \\ \hdashline[3pt/2pt]
\multirow{2}{*}{\textnormal{\footnotesize \#Bedrooms}} & \textnormal{\footnotesize 0} & \textnormal{\footnotesize 1} & \textnormal{\footnotesize 2} & \textnormal{\footnotesize 3} & \textnormal{\footnotesize 4} & \textnormal{\footnotesize 5} & \textnormal{\footnotesize More than 5} \\
& \textnormal{\footnotesize 0\%} & \textnormal{\footnotesize 14\%} & \textnormal{\footnotesize 28\%} & \textnormal{\footnotesize 41\%} & \textnormal{\footnotesize 14\%} & \textnormal{\footnotesize 3\%} & \textnormal{\footnotesize 0\%} \\ \hdashline[3pt/2pt]
\textnormal{\footnotesize \#Rooms other} & \textnormal{\footnotesize 0} & \textnormal{\footnotesize 1} & \textnormal{\footnotesize 2} & \textnormal{\footnotesize 3} & \textnormal{\footnotesize 4} & \textnormal{\footnotesize 5} & \textnormal{\footnotesize More than 5} \\
\textnormal{\footnotesize than bedrooms}& \textnormal{\footnotesize 4\%} & \textnormal{\footnotesize 11\%} & \textnormal{\footnotesize 17\%} & \textnormal{\footnotesize 27\%} & \textnormal{\footnotesize 26\%} & \textnormal{\footnotesize 15\%} & \textnormal{\footnotesize 0\%} \\
\bottomrule
\end{tabular}
\caption{Breakdown of participants' home settings.}
\label{tab:home_settings}
\end{table*}

\begin{table*}[h]
\centering
\setlength{\tabcolsep}{4pt}
\begin{tabular}{lrlrlrlrlrlr}
\toprule
{\footnotesize Age} & & \multicolumn{2}{l}{\footnotesize Gender} & 
\multicolumn{2}{l}{\footnotesize Highest Degree} & \multicolumn{2}{l}{\footnotesize Country of Residence} & \multicolumn{2}{l}{\footnotesize Employment} & \multicolumn{2}{l}{\footnotesize Tech Background} \\ \hline
\textnormal{\footnotesize 18-29} & \textnormal{\footnotesize 62\%} & \textnormal{\footnotesize Female} & \textnormal{\footnotesize 43\%} & \textnormal{\footnotesize No schooling completed} & \textnormal{\footnotesize 0\%} & \textnormal{\footnotesize UK} & \textnormal{\footnotesize 31\%} & \textnormal{\footnotesize Full-time} & \textnormal{\footnotesize 41\%} & \textnormal{\footnotesize Yes} & \textnormal{\footnotesize 35\%} \\
\textnormal{\footnotesize 30-49} & \textnormal{\footnotesize 34\%} & \textnormal{\footnotesize Male} & \textnormal{\footnotesize 57\%} & \textnormal{\footnotesize Nursery school} & \textnormal{\footnotesize 0\%} & \textnormal{\footnotesize Poland} & \textnormal{\footnotesize 15\%} & \textnormal{\footnotesize Part-time} & \textnormal{\footnotesize 17\%} & \textnormal{\footnotesize No} & \textnormal{\footnotesize 65\%} \\
\textnormal{\footnotesize 50-64} & \textnormal{\footnotesize 4\%} &  &  & \textnormal{\footnotesize Grades 1 through 11} & \textnormal{\footnotesize 2\%} & \textnormal{\footnotesize US} & \textnormal{\footnotesize 14\%} & \textnormal{\footnotesize Unemployed} & \textnormal{\footnotesize 6\%} & &  \\
& &  &  & \textnormal{\footnotesize 12\textsuperscript{th} grade—no diploma} & \textnormal{\footnotesize 3\%} & \textnormal{\footnotesize Italy} & \textnormal{\footnotesize 7\%} & \textnormal{\footnotesize Self-employed} & \textnormal{\footnotesize 8\%} & & \\
& &  &  & \textnormal{\footnotesize Regular high-school diploma} & \textnormal{\footnotesize 21\%} & \textnormal{\footnotesize Portugal} & \textnormal{\footnotesize 7\%} & \textnormal{\footnotesize Home-maker} & \textnormal{\footnotesize 3\%} & & \\
& &  &  & \textnormal{\footnotesize GED or alternative credential} & \textnormal{\footnotesize 0\%} & \textnormal{\footnotesize Spain} & \textnormal{\footnotesize 4\%} & \textnormal{\footnotesize Student} & \textnormal{\footnotesize 25\%} & & \\
& &  &  & \textnormal{\footnotesize Some college credit, $<1$ year of college} & \textnormal{\footnotesize 4\%} & \textnormal{\footnotesize Greece} & \textnormal{\footnotesize 3\%} & \textnormal{\footnotesize Retired} & \textnormal{\footnotesize 0\%} & & \\ \cdashline{9-10}
& &  &  & \textnormal{\footnotesize 1+ years of college credit, no degree} & \textnormal{\footnotesize 16\%} & \textnormal{\footnotesize Canada} & \textnormal{\footnotesize 2\%} &  \textnormal{\footnotesize Public sector} & \textnormal{\footnotesize 23\%}  & & \\
& &  &  & \textnormal{\footnotesize Associate's degree (e.g., AA, AS)} & \textnormal{\footnotesize 4\%} & \textnormal{\footnotesize Other} & \textnormal{\footnotesize 17\%} &  \textnormal{\footnotesize Private sector} & \textnormal{\footnotesize 67\%} & & \\
& &  &  & \textnormal{\footnotesize Bachelor's degree (e.g., BA, BS)} & \textnormal{\footnotesize 37\%} & & &  \textnormal{\footnotesize Non-profit sector} & \textnormal{\footnotesize 10\%} & & \\
& &  &  & \textnormal{\footnotesize Master's degree (e.g., MA, MS, MBA)} & \textnormal{\footnotesize 13\%} & & &  &  & & \\
& &  & & \textnormal{\footnotesize Professional degree (e.g., MD, JD)} & \textnormal{\footnotesize 0\%} &  &  & & \\
& &  &  & \textnormal{\footnotesize Doctorate degree (e.g., Ph.D., EdD)} & \textnormal{\footnotesize 0\%} &  &  &  & & \\
\bottomrule
\end{tabular}
\caption{Participants' demographic information. Only countries with at least 5 participants are listed.}
\label{tab:demographics}
\end{table*}

\end{document}